\documentclass{aa}

\newcommand{\lya}{\mbox{${\rm Ly}\alpha$}}

\usepackage{graphicx,amsmath,amssymb,epsfig}
\usepackage[T1]{fontenc}
\usepackage{multirow}
\usepackage{color}
\usepackage{hyperref}

\begin{document}

\title{Dissecting cold gas in a high-redshift galaxy using a lensed background quasar}

\author{
J.-K. Krogager\inst{1} \and
P. Noterdaeme\inst{1} \and
J. M. O'Meara\inst{2} \and
M. Fumagalli\inst{3,4} \and
J. P. U. Fynbo\inst{5} \and
J. X. Prochaska\inst{6,7} \and
J. Hennawi\inst{8,9} \and
S. Balashev\inst{10} \and
F. Courbin\inst{11} \and
M. Rafelski\inst{12,13} \and
A. Smette\inst{14} \and
P. Boiss\'e\inst{1}
}

\institute{
	Institut d'Astrophysique de Paris, CNRS-SU, UMR7095, 98bis bd Arago, 75014 Paris, France\\
	\email{jens-kristian.krogager@iap.fr}
\and
	Department of Chemistry and Physics, Saint Michael's College, One Winooski Park, Colchester, VT 05439
\and
	Centre for Extragalactic Astronomy, Durham University, South Road, Durham, DH1 3LE, UK
\and
	Institute for Computational Cosmology, Durham University, South Road, Durham, DH1 3LE, United Kingdom
\and	
	The Cosmic Dawn Center, Niels Bohr Institute, University of Copenhagen,\\
	Juliane Maries Vej 30, 2100 Copenhagen \O, Denmark
\and
	Department of Astronomy and Astrophysics, University of California,\\
	1156 High Street, Santa Cruz, CA 95064, USA
\and
	University of California Observatories, Lick Observatory,
	1156 High Street, Santa Cruz, CA 95064, USA
\and
	Department of Physics, Broida Hall, University of California, Santa Barbara, CA 93106, USA
\and
	Max-Planck-Institut f\"ur Astronomie, K\"onigstuhl 17, D-69117 Heidelberg, Germany
\and
	Ioffe Institute, Polytechnicheskaya ul. 26, Saint Petersburg, 194021, Russia
\and
	Institute of Physics, Laboratory of Astrophysics, \'Ecole Polytechnique F\'ed\'erale de Lausanne (EPFL),
	Observatoire de Sauverny, 1290, Versoix, Switzerland
\and
	Space Telescope Science Institute, 3700 San Martin Drive, Baltimore, MD 21218, USA
\and
	Department of Physics \& Astronomy, Johns Hopkins University, Baltimore, MD 21218, USA
\and
	European Southern Observatory, Alonso de C\'ordova 3107, Vitacura, Santiago, Chile
}

\abstract{
	We present a study of cold gas absorption from a damped Lyman-$\alpha$ absorber (DLA) at
	redshift $z_{\rm abs}=1.946$ towards two lensed images of the quasar J144254.78+405535.5
	at redshift $z_{\rm \textsc{qso}} = 2.590$. The physical separation of the two lines of
	sight at the absorber redshift is $d_{\rm abs}=0.7$~kpc based on our lens model.
	We observe absorption lines from neutral carbon and H$_2$ along both lines of sight
	indicating that cold gas is present on scales larger than $d_{\rm abs}$.
	We measure column densities of \ion{H}{i} to be $\log N(\rm H\,\textsc{i}) = 20.27\pm0.02$
	and $20.34\pm0.05$ and of H$_2$ to be $\log N(\rm H_2) = 19.7\pm0.1$ and $19.9\pm0.2$.
	The metallicity inferred from sulphur is consistent with Solar metallicity for both sightlines:
	$[{\rm S/H}]_A = 0.0\pm0.1$ and $[{\rm S/H}]_B = -0.1\pm0.1$.
	Based on the excitation of low rotational levels of H$_2$, we constrain the temperature
	of the cold gas phase to be $T=109\pm20$ and $T=89\pm25$~K for the two lines of sight.
	From the relative excitation of fine-structure levels of \ion{C}{i}, we constrain the hydrogen
	volumetric densities in the range of $40-110$~cm$^{-3}$. Based on the ratio of observed
	column density and volumetric density, we infer the average individual `cloud' size along
	the line of sight to be $l\approx0.1$~pc. Using the transverse line-of-sight separation
	of 0.7~kpc together with the individual cloud size, we are able to put an upper limit
	to the volume filling factor of cold gas of $f_{\rm vol} < 0.2$~\%.
	Nonetheless, the projected covering fraction of cold gas must be large (close to unity)
	over scales of a few kpc in order to explain the presence of cold gas in both lines of sight.
	Compared to the typical extent of DLAs ($\sim10-30$~kpc), this is consistent with the
	relative incidence rate of \ion{C}{i} absorbers and DLAs.
}


\titlerunning{Cold gas at high-$z$ towards gravitationally lensed quasar}
\authorrunning{Krogager et al.}

\maketitle

% ----------------------  Intro  -------------------------------------
\section{Introduction}

The onset of star formation is intimately linked to the cooling and subsequent collapse of the neutral gas in galaxies. Under pressure equilibrium, the neutral gas naturally segregates into two distinct temperature phases: a cold ($T\sim100$~K) neutral medium (CNM) and a warm ($T\sim10^3-10^4$~K) neutral medium (WNM) as characterised locally in the canonical two-phase model \citep[e.g.,][]{Field1969, Wolfire1995}.
In the distant Universe, the neutral gas phase is most readily accessible through observations of damped \lya\ absorbers \citep[DLAs;][]{Wolfe1986, Barnes2014_review}, which make up the class of the highest column density \lya\ absorbers defined as having $N_{\rm H\,\textsc{i}} > 2 \times 10^{20}$~cm$^{-2}$. However, as the presence and strength of \lya\ absorption does not depend on temperature, the \lya\ line alone does not constrain the relative contribution of CNM and WNM in DLAs.
Since the interplay between the warm and cold neutral gas is crucial for the regulation of star formation, understanding how DLAs trace the CNM and WNM is therefore of great importance for galaxy evolution studies.

A direct way of probing the cold neutral gas at high redshift is through the use of \ion{H}{i} 21-cm absorption studies since its optical depth does depend on temperature \citep{Gupta2009, Gupta2012, Curran2010, Srianand2012, Kanekar2014, Dutta2017}. 
Based on the spin-temperature measurements from \ion{H}{i} 21-cm absorption, \citet{Kanekar2014} find that the mass fraction of gas in DLAs with similar characteristics as the locally observed CNM must be $\lesssim$20\%, indicating that the CNM fraction is lower at high-redshift than what is seen locally.

Another method of probing the physical conditions in DLAs is through the use of excited fine-structure levels of metal absorption lines, e.g., \ion{C}{i}, \ion{C}{ii} and \ion{Si}{ii} \citep[e.g.,][]{Wolfe2003a, Wolfe2008, Howk2005, Srianand2005, Jorgenson2010, Neeleman2015}. Since these fine-structure transitions are mainly excited via collisions, the population ratio depends on density and temperature of the absorbing gas. For a large sample of DLAs, \citet{Neeleman2015} find that about 5\% show significant amounts of cold gas absorption.
For the remaining absorbers, the physical properties are not well constrained due to limitations in the modelling of fine-structure levels. Yet the constraints are consistent with the gas being warm similar to the metal-poor DLAs studied by \citet{Cooke2015}, who infer low densities ($n\sim0.1$~cm$^{-3}$) and high temperatures ($T\sim5000$~K). Similar temperatures of $T\sim10^4$~K for the warm gas phase are inferred directly from thermal Doppler broadening of ions with different masses \citep{Carswell2012, Noterdaeme2012b}.

Molecular hydrogen provides a very useful and direct tracer of the CNM since its presence requires the gas to be cold and shielded from radiation \citep{Krumholz2009, Krumholz2012}. Nevertheless, the Lyman and Werner absorption bands in the rest-frame ultraviolet fall in the \lya\ forest and it can be extremely challenging to detect these absorption lines especially at low spectral resolution \citep{Balashev2014}.
Targeted studies of H$_2$ absorption for pre-selected DLAs show that the fraction of DLAs with detectable amounts of H$_2$ is very low, typically $\lesssim10$~\% \citep{Petitjean2000, Ledoux2003, Noterdaeme2008, Jorgenson2014, Balashev2018}.

Instead of relying on the H$_2$ absorption lines as the main tracer of the CNM gas in DLAs, we can use absorption from \ion{C}{i} as another proxy for the cold gas \citep{Srianand2005, Jorgenson2010, Noterdaeme2018}. Since the ionization potential of \ion{C}{i} (11.3~eV) is lower than that of \ion{H}{i} (13.6~eV), neutral carbon absorption arises only in highly shielded regions where the gas is able to cool efficiently. This makes \ion{C}{i} an excellent tracer of the CNM (albeit limited to high-metallicity systems) and provides a direct probe of the cold gas phase. Moreover, the rest-frame transitions of the strongest absorption features from \ion{C}{i} typically fall outside the \lya\ forest thereby providing the basis for an efficient selection of the CNM at high redshift \citep{Ledoux2015}.

While we can now directly trace the cold gas phase using proxies such as \ion{C}{i}-absorption, it is still not possible to obtain meaningful constraints on the filling factor of the cold gas since we only have one line of sight through the medium when using quasar absorbers. In a handful of cases, we are able to detect absorption systems in two lines of sight using closely projected quasar pairs \citep{Hennawi2006, Hennawi2010, Prochaska2013, Rubin2015}. However, such projected pairs typically probe large separations $d \gtrsim 10$~kpc, and in no cases has \ion{C}{i} gas been observed in both sightlines.

If instead the background source is a strongly lensed quasar, the separation between the multiple lines of sight will change as a function of redshift, reaching a maximum separation at the lens redshift and converging at the source redshift. This provides a unique way of probing smaller physical separations ($\lesssim5$~kpc) if the absorption system is located between the source and the lens \citep[e.g.,][]{Smette1995, Michalitsianos1997, Churchill2003, Ellison2007, Cooke2010}. While this has resulted in interesting constraints on the extent of the neutral gas phase (as probed by \ion{H}{i}), the cold gas properties on small scales remain unconstrained.

At lower redshifts ($z<1$), three molecular, intervening absorption systems are known towards lensed radio quasars \citep{Wiklind1995, Wiklind1996, Kanekar2005}. In these cases, the absorption arises in the lens galaxies themselves and probe typical line-of-sight separations of $\sim5$~kpc. The absorption lines detected at radio or sub-mm wavelengths allow a detailed study of the kinematics and physical conditions in the lens galaxies. However, studying absorption in lensing galaxies at higher redshifts ($z \gtrsim 2$) becomes increasingly difficult as high-redshift lens galaxies are extremely rare \citep{Oguri2010}.

In this paper, we report the first detection of a high-redshift intervening absorber towards a lensed quasar at redshift $z_{\rm em}=2.59$ showing \ion{C}{i} absorption in both lines of sight at redshift $z_{\rm abs}=1.946$. The absorber is not associated to the lensing galaxy as such a configuration would be unphysical given the quasar redshift and the observed image separation (2.1~arcsec). The primary image (also the brightest, hereafter image A) of the quasar J144254.78+405535.5 (hereafter J1442+4055) was selected through a direct search of \ion{C}{i} absorbers in the Sloan Digital Sky Survey \citep[as in the work by][]{Ledoux2015}. Through serendipitous inspection of the field, a nearby, unidentified point source with similar colours was targeted as a quasar pair candidate. Subsequent spectroscopic follow-up revealed the nearby point source to be a lensed image of the same quasar (hereafter image B). The quasar lens was simultaneously, yet independently, discovered by a targeted search for quasar lens candidates \citep{More2016, Sergeyev2016}.

Due to the absorbers position in between the lens and the quasar, the line-of-sight separation at $z=1.946$ is smaller than the maximal, projected separation in the lens plane. We are therefore able to study the cold gas in this absorber over small transverse scales ($\approx1$~kpc). If the two sources had been a close pair instead of a lensed quasar, the physical separation at the absorber redshift would have been $\sim17$~kpc instead. The small physical separation between the lines of sight enables us, for the first time, to put direct constraints on the volume filling factor of cold gas in a high-redshift, absorption-selected galaxy.

The paper is structured as follows: in Sect.~\ref{observations}, we describe the observations and data processing; in Sect.~\ref{lens_model}, we describe the lens model in order to obtain the line-of-sight separation at the absorber redshift; in Sect.~\ref{results}, we describe the data analysis and measurements; and in Sect.~\ref{discussion}, we present the interpretation of the derived observables and discuss our findings.
Throughout this paper, we will assume a flat $\Lambda$CDM cosmology with $\Omega_{\Lambda} = 0.7$ and $H_0 = 68$~km~s$^{-1}$~Mpc$^{-1}$ \citep{Planck2014}.

\section{Observations and Data Processing}
\label{observations}

The spectroscopic identification of the second quasar image J144254.60+405535.0 (hereafter J1442+4055B) was secured using the ALFOSC spectrograph mounted on the Nordic Optical Telescope (NOT) at Observatorio Roque de los Muchachos, La Palma, Spain through the Fast-Track programme (observing ID: P52-415). The low resolution spectroscopy using ALFOSC was obtained on Apr 28, 2016 using grism \#4 covering the whole optical wavelength range (3200$-$9600~\AA) with a spectral resolution of roughly $R=360$. The spectrum was taken using a slit-width of 1.3~arcsec during good conditions with a seeing of 1.0~arcsec.

\begin{figure*}
    \centering
    \includegraphics[width=0.99\textwidth]{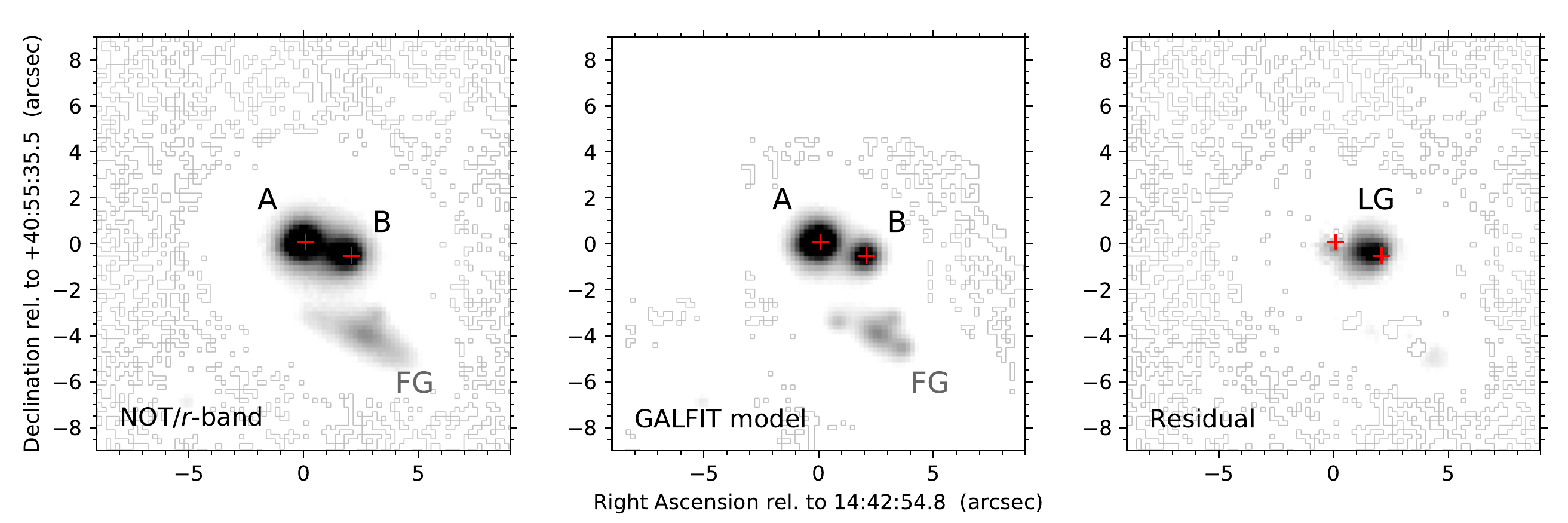}
    \caption{Imaging of the field around the lensed quasar in $r$-band (left).
	In the middle panel, we show a model of the two quasar images (marked A and B) and an
	unrelated foreground galaxy (FG) using GALFIT \citep{Peng2002}. The centroids of the two
	quasar images are shown by red crosses in all frames.
	The right panel shows the residuals of the model-subtracted image, and
	the lensing galaxy (LG) is clearly visible. The same color scale is used for all frames.}
    \label{fig:imaging}
\end{figure*}

In order to search for emission from the lensing galaxy, we observed the field using the instrument ALFOSC in imaging mode to obtain a $r$-band image. The observations were carried out on Aug 3, 2016 using a 9-point dither pattern with individual integration times of 270 sec resulting in a total integration time of 2430 sec. The images were bias subtracted using a median combined bias frame (based on 11 individual bias frames), and flat field correction was performed using sky-flats obtained in twilight. Afterwards, the images were background subtracted using the median sky value, and all images were subsequently shifted and combined into a single frame. The combined $r$-band image is shown in Fig.~\ref{fig:imaging}.
We construct a model for the two quasar images using {\sc Galfit} \citep{Peng2002} assuming that they are unresolved point sources using a combined point spread function from combined stars in the field. Moreover, we model the foreground galaxy located south of the lens images using a S\'ersic profile \citep{Sersic1963}. We note that the lensing galaxy is included in the fit in order not to bias the fit of the two quasar images. However, the model shown in Fig.~\ref{fig:imaging} only includes the two quasar images and the foreground galaxy in order to highlight the lensing galaxy in the residual image (right-hand panel of Fig.~\ref{fig:imaging}). The lensing galaxy is clearly visible after subtracting the two point sources \citep[see also][]{Sergeyev2016}.

The quasar lens was observed at low resolution using the LRIS spectrograph \citep{Oke1995}
at the Cassegrain focus of Keck-I on June 5, 2016.  A slit-width of 1~arcsec was used,
and the slit was oriented to cover both members of the pair simultaneously.
The 560 dichroic, 600/7500 (0.8~\AA/pixel) grating, and 600/4000 (0.6~\AA/pixel) grisms
were used to provide continuous wavelength coverage from $3090-8651$~\AA.
The data were flat fielded, optimally extracted, and flux calibrated
(using the Feige 34 spectrophotometric standard star) with the
LowRedux\footnote{\url{http://www.ucolick.org/~xavier/LowRedux/}} code.

The two quasar images were observed at high resolution using the HIRES spectrograph \citep{Vogt1994}
on May 20, 2017. Each observation was made using the C1 decker, which provides a spectral resolution
of R$\sim$48\,000. The slit was aligned to have only one member of the quasar images in the slit at
any one time. The choice of decker was made in part to further suppress contamination of one image
by the other. The data span the wavelength range of $3029 < \lambda < 5880$~\AA.
The quasar A was observed for at total of 5600 seconds over three integrations,
and the quasar B was observed for 7200 seconds over three integrations.

The data were reduced using the HIRedux\footnote{\url{http://www.ucolick.org/~xavier/HIRedux/}} code
which is a part of the XIDL\footnote{\url{http://www.ucolick.org/~xavier/IDL/index.html}} package of
astronomical routines. 
The data reduction and continuum fitting were performed in the same manner as presented in \citet{OMeara2015}.

\subsection{Archival SDSS Imaging}

In order to constrain the lensing galaxy in greater detail, we use the archival SDSS imaging data of the system
in the 5 SDSS filters ($u$, $g$, $r$, $i$, and $z$). We perform a similar modelling as described above for the NOT data using {\sc Galfit} to isolate the lensing galaxy. The lensing galaxy is undetected in the $u$-band, marginally detected in the $g$-band, but clearly detected in the $r$, $i$ and $z$ bands.
However, since the image quality is worse than our deep NOT $r$-band, we use the structural parameters derived from the NOT data to constrain the fit to the SDSS data, i.e., we keep the S\'ersic index fixed to $n=4$, the axis-ratio is fixed to 0.9, and the half-light radius is fixed to $r_e=3.2$~pixels. For the $g$-band, we use a slightly lower index of $n=3$ and larger $r_e=4$~pixels as galaxies tend to have lower indices and larger extent in the bluer rest-frame filters \citep{Kelvin2012}.
The derived magnitudes in the detected filters are: $g=21.0\pm0.6$, $r=19.3\pm0.1$, $i=18.7\pm0.1$, and $z=18.1\pm0.2$ all on the AB system. For comparison, the fainter quasar image B is similar in apparent magnitude having an $r$-band magnitude of $r=19.0$ whereas image A is significantly brighter with $r=18.3$.

\section{Lens Model}
\label{lens_model}

\subsection{Lensing Galaxy Properties}

Based on the photometry from the SDSS images, we derive a photometric redshift estimate for the lensing galaxy using {\sc Eazy} \citep{Brammer2008}. The derived photometric redshift is $z_{\rm phot} = 0.35\pm0.15$ based on the median of the posterior distribution. This estimate is highly uncertain due to the poor photometry and the low number of filters available, yet it serves as a good initial guess for further analysis.

From Fig.~\ref{fig:imaging}, we see that the lensing galaxy is located closer to quasar image B than to image A.
Hence, from the two LRIS spectra of image A and B where the slit was aligned to have both images in the slit simultaneously, we expect the lensing galaxy to contribute more to the spectrum of image B than spectrum of image A.
We can quantify this expected contribution from the galaxy flux in the two spectra by using the NOT $r$-band image, since the NOT and LRIS observations were taken under similar seeing conditions.
By imitating the slit orientation on top of the imaging data, we can calculate the amount of flux that would pass through a 1~arcsec slit. Applying the slit aperture to the $r$-band image image results in a 1-dimensional spatial profile consistent with what is observed in the LRIS data. Furthermore, since we have decomposed the $r$-band image into two quasar image contributions and lensing galaxy contribution, we can calculate the amount of galaxy flux expected in each spectrum. For spectrum A, the fraction of the total lensing galaxy flux is roughly 8~\% whereas for spectrum B the fraction of the total lensing galaxy flux is 32~\%.

Having constrained the relative amount of galaxy flux in the two spectra, we can fit a combined quasar and lens galaxy model to the two spectra. For this purpose, we use the quasar template by \citet{Selsing2016} as a model for the intrinsic spectral shape of the quasar images and as a description of the lensing galaxy, we use the set of galaxy templates from {\sc Eazy} generated with {\sc P\'egase} \citep[][see \citet{Brammer2008} for details describing the templates]{Fioc1997}. Since {\sc Eazy} has already provided a best-fit template when estimating the photometric redshift, we use this template as an initial guess.
Since the cold-gas absorber at $z_{\rm abs} = 1.946$ could potentially harbour dust which would redden the background quasars, we include a dust contribution at the absorber rest-frame to the model.
We do not consider any contribution to the reddening of the quasar images from the lensing galaxy itself since we do not observe any significant absorption features from \ion{Na}{i} and \ion{Ca}{ii} in the redshift range constrained from the posterior of $z_{\rm phot}$. As the strength of \ion{Na}{i} and \ion{Ca}{ii} is observed to correlate with dust reddening \citep{Wild2005, Murga2015}, we rule out any significant reddening from the lensing galaxy at the projected position of the quasar images.
Our assumption that the dust arises only in the high-redshift absorber is further bolstered by the fact that the best-fit values of A(V) are consistent with the inferred A(V) based on the depletion and metallicity (see Sect.~\ref{metallicity}).

The combined model for spectrum A (and similarly for B) is then given as:
\begin{equation}
	f_A(\lambda) = r_A \ T_{\rm QSO}\,(\lambda) \ 10^{-0.4{\xi}_A(\lambda)A(V)_A} + x_A \ r_G \ T_{\rm gal}\,(\lambda, z_L)~,
\end{equation}

\noindent
where $r_A$ denotes the absolute flux scaling of the quasar template, $T_{\rm QSO}$, $\xi_A$ denotes the reddening law in the rest-frame of the absorber towards quasar image A, ${\rm A(V)}_A$ denotes the rest-frame $V$-band extinction, $x_A$ is the fraction of the total galaxy flux in spectrum A, and $r_G$ is the absolute flux scaling of the galaxy template, $T_{\rm gal}$, at lens redshift $z_L$. Since the reddening law might be different along the two lines of sight, we include two separate terms ${\xi}_A$ and ${\xi}_B$ in the model. The two factors $x_A=0.08$ and $x_B=0.32$ are constrained from the imaging analysis. We fit both spectra and their ratio, $f_A / f_B$, simultaneously (as the ratio of the two adds additional constraints on the reddening and the galaxy spectrum).

We find that in both cases the spectra are well reproduced using the reddening law inferred for the Small Magellanic Cloud (SMC), however, with a stronger 2175~\AA\ bump along sightline B. We parametrize the reddening law using the formalism of \citet{FM2007} keeping the parameters fixed to those of the average SMC reddening law except for the parameter $c_3$ for sightline B which controls the strength of the 2175~\AA\ bump.

Initially, we fix the reddening law parameters and vary only ${\rm A(V)}_A$, ${\rm A(V)}_B$, $r_A$, $r_B$, $r_G$, and $z_L$. This gives us a first estimate of the lens galaxy spectrum for the assumed galaxy template. We then fit the whole library of galaxy templates to the isolated lens galaxy spectrum in order to optimize the galaxy template. After having converged on a galaxy template (of a single stellar population with ages of $\sim$11~Gyr) we fit the set of parameters again, this time varying also the bump strength along sightline B. Since the parameters are highly degenerate and the spectra might be affected by additional microlensing, we were not able to constrain the bump strength independently. Hence, we fixed the bump strength at $c_3=1.5$ for sightline B which corresponds to the bump strength observed for the LMC2 reddening law. This value of $c_3$ provides the optimal lens galaxy spectrum.

\begin{figure}
	\includegraphics[width=1.0\columnwidth]{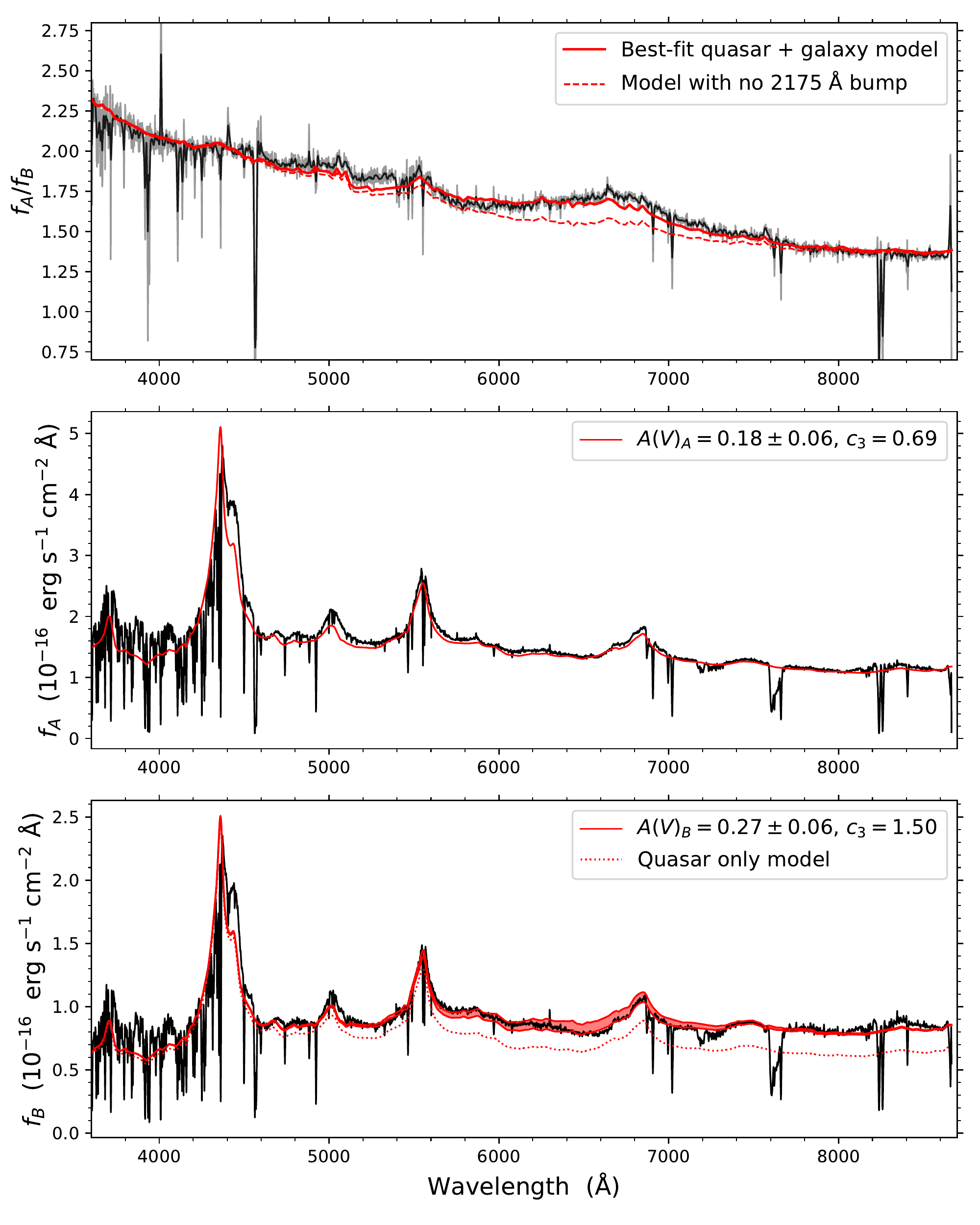}
	\caption{Spectra for the two lines of sight using the Keck/LRIS data.
	In the top panel, we show the ratio of the two spectra together with the
	best-fit quasar and galaxy model in red (see text). For comparison,
	we show the same model with no 2175~\AA\ bump.
	In the middle and bottom panels, we show the individual spectra
	together with their best-fit dust-reddened quasar plus galaxy model in red.
	The shaded region in the bottom panel indicates the location and strength of the
	2175~\AA\ dust bump. The dotted line in the bottom panel shows the quasar-only
	contribution to the model.
	\label{fig:reddening}
	}
\end{figure}

\begin{figure}
	\includegraphics[width=1.0\columnwidth]{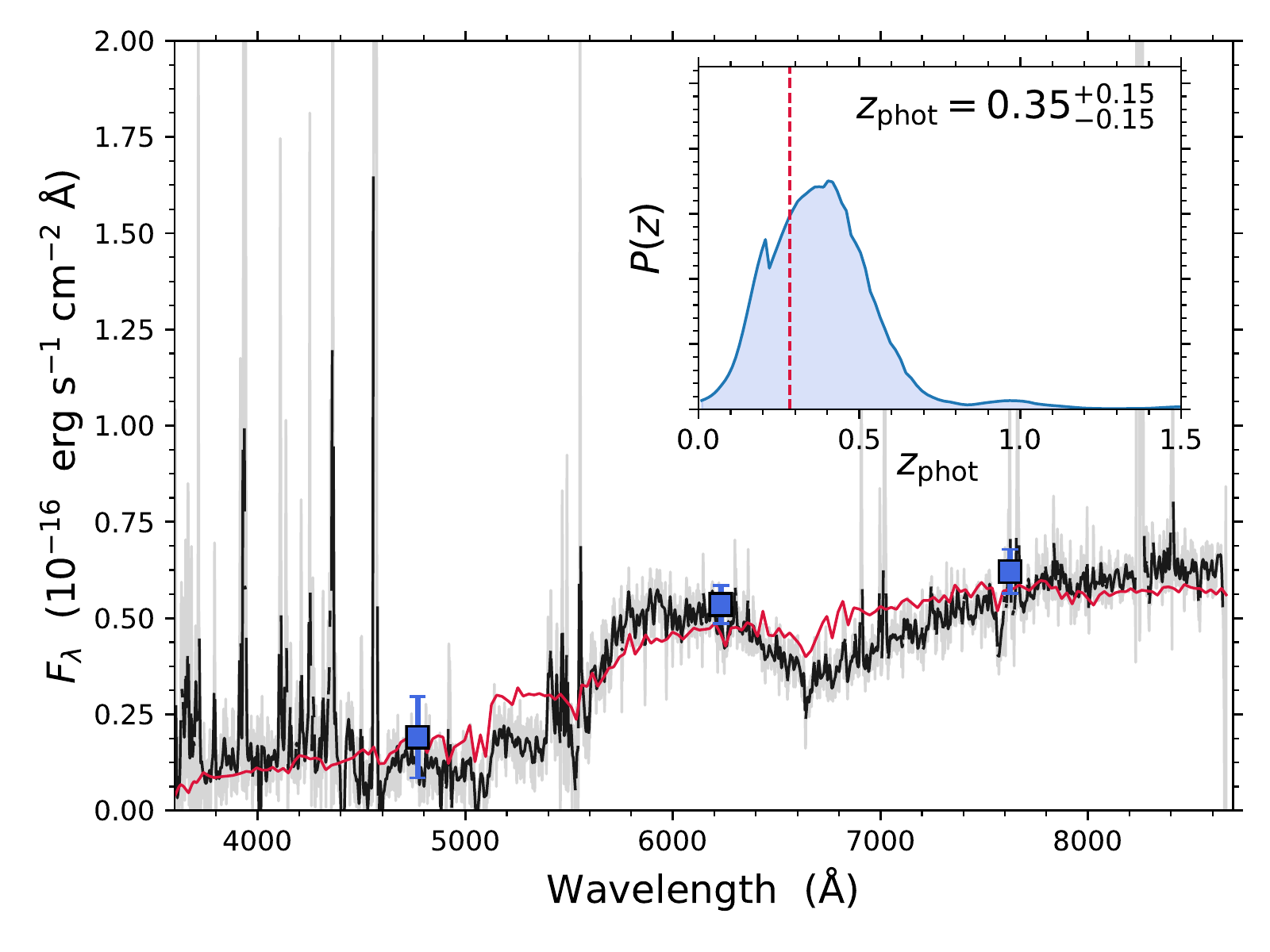}
	\caption{Isolated lens galaxy spectrum recovered from spectral fitting. The light gray line
	shows the raw spectrum and in black we show the Gaussian filtered spectrum with a median clipping
	to remove outlying pixels caused by variations in absorption lines from the $z_{\rm abs}=1.946$ system.
	The red line indicates the best-fit galaxy template of a single stellar population ($t=11$~Gyr),
	and the blue squares show the SDSS photometry of the lensing galaxy.
	The inset in the upper right corner shows the posterior probability of the photometric redshift
	estimation from {\sc Eazy}, where the red, dashed line marks the best-fit lens redshift.
	\label{fig:lens_spectrum}
	}
\end{figure}

The best-fit solution was obtained for the following parameters: ${\rm A(V)}_A = 0.18 \pm 0.06$~mag, ${\rm A(V)}_B = 0.27 \pm 0.06$~mag, and $z_L=0.284\pm0.001$.
The uncertainty on A(V) is here dominated by the systematic uncertainty related to the unknown intrinsic shape of the quasar spectrum. From the observed dispersion of quasar power-law shapes \citep{Krawczyk2015}, we obtain the systematic 1$\sigma$ uncertainty of $0.06$~mag.
The LRIS data together with the best-fit model are shown in Fig.~\ref{fig:reddening}, and the resulting isolated lens galaxy spectrum together with the optimal template is shown in Fig.~\ref{fig:lens_spectrum}. The lensing galaxy spectrum and the optimal template have been corrected for slit-loss\footnote{Based on the simulated slit aperture of the imaging data, we calculate that only 40\% of the galaxy flux falls within the 1~arcsec slit, of which 80\% is included in the extraction aperture of spectrum B.}. In Fig.~\ref{fig:lens_spectrum} we also show the SDSS photometry in $g$, $r$ and $i$. Due to the galaxy being an extended source, only 40~\% of the total flux makes it into 1~arcsec slit. Overall the re-scaled photometry is fully consistent with the inferred lens galaxy spectrum.

The spectral region from $\sim5000-7000$~\AA\ is possibly affected by variations in the broad emission lines together with the iron pseudo-continuum from a blend of many \ion{Fe}{ii} and \ion{Fe}{iii} emission lines. Nonetheless, the absorption features from \ion{Ca}{ii}, \ion{Mg}{i}, and \ion{Na}{i} lines as well as the 4000~\AA\ break in the galaxy spectrum strongly constrain the lens redshift.

The A(V) measurements are not significantly affected by the bump strength (the variations are around 0.01~mag, less than the systematic uncertainty due to quasar continuum shape), but we note that the reddening law cannot be constrained independently since the shape of the reddening law is degenerate with the galaxy and quasar template shapes. This is further complicated by a possible microlensing effect of the quasar continuum.
Although microlensing in general is achromatic, it can lead to chromatic effects if the continuum emitting region has a radial colour gradient as is expected for most accretion disk models \citep[e.g.,][]{Wambsganss1991}. However, with only one epoch of observations those effects are impossible to disentangle.

\subsection{Line-of-sight Separation}

\begin{figure}
    \centering
    \includegraphics[width=0.5\textwidth]{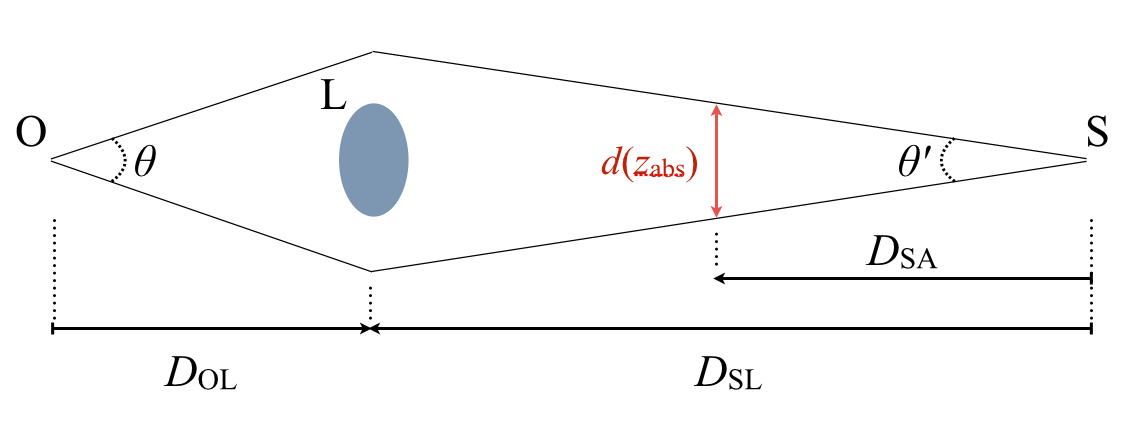}
    \caption{The lensing geometry. The distances between observer (O), lens (L),
		source (S), and absorber (A) at redshift $z_{\rm abs}$ are illustrated.}
    \label{fig:geometry}
\end{figure}

We calculate the distance between the two quasar sightlines
as a function of redshift from $z=0$ to the redshift of the quasar following
\citet{Smette1992}.
A simplified geometry is illustrated in Fig.~\ref{fig:geometry}.

The angle on the sky between the two images is denoted $\theta$, the distance between the two sightlines is denoted $d(z)$ and the angular diameter distances between source and lens, source and absorber, and observer and lens are denoted $D_{SL}$, $D_{SA}$ and $D_{OL}$, respectively. From \citet{Smette1992}, we calculate $d(z)$ as:

\begin{equation}
	d(z) = \theta \frac{D_{SA} D_{OL}}{D_{SL}}~.
\end{equation}

\begin{figure}
    \centering
    \includegraphics[width=0.48\textwidth]{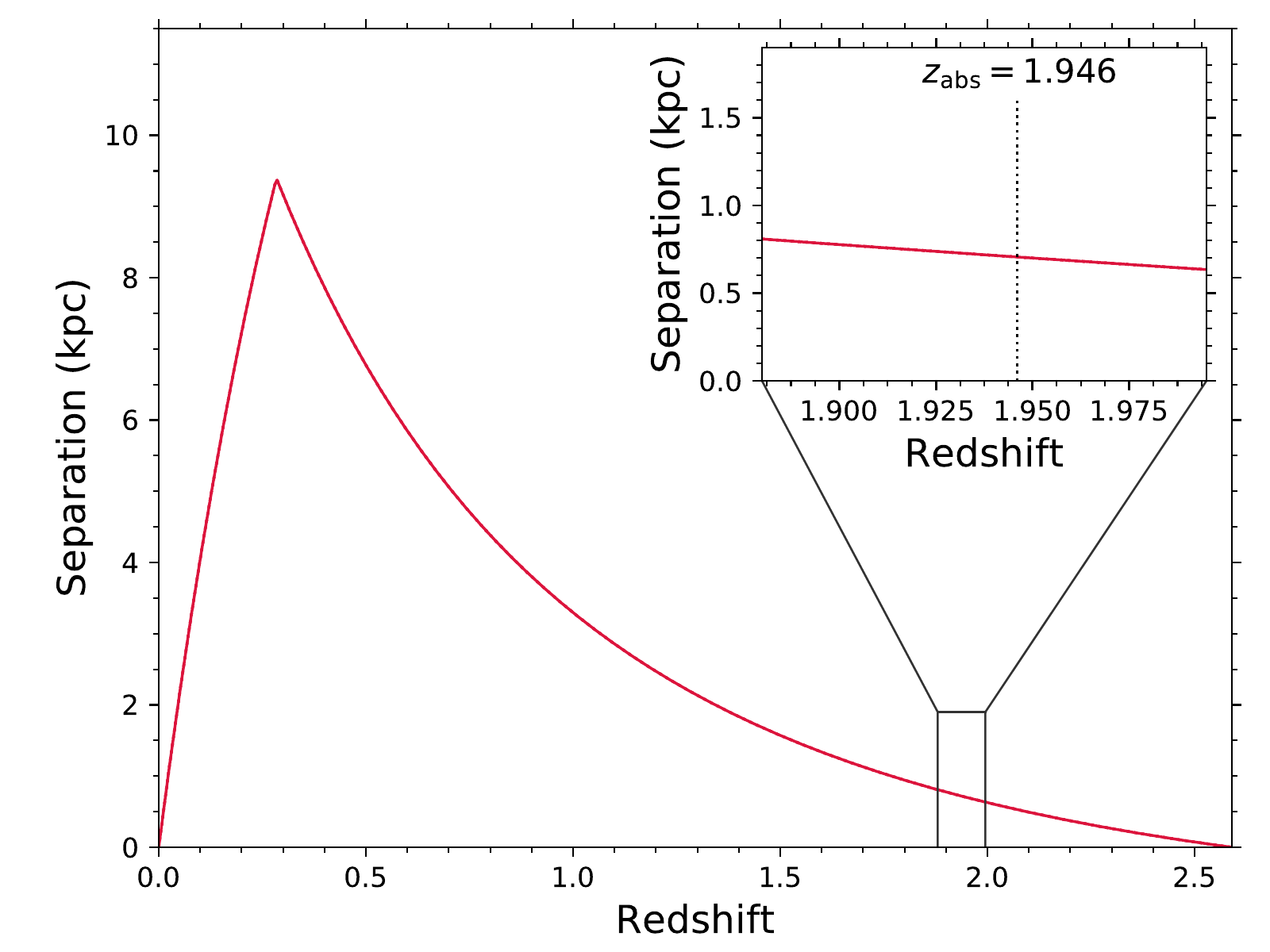}
    \caption{The distance between the two sightlines as a function of
		redshift assuming a lens redshift of $z_L = 0.284$.}
    \label{fig:distance}
\end{figure}

We then calculate the line-of-sight separation as a function of redshift
for the case of J1442+4055 assuming a lens redshift of $z_L = 0.284$
and a source redshift of $z_S = 2.590$.
From imaging of the field obtained at the Nordic Optical Telescope (see Fig.~\ref{fig:imaging}),
we find $\theta = 2.13\pm0.01$~arcsec.
The resulting line-of-sight separation as a function of redshift is shown in Fig.~\ref{fig:distance},
and at the redshift of the absorber
$z_{\rm abs} = 1.946$ we find $d_{\rm abs} = 0.71\pm0.01$~kpc for the best-fit lens redshift of $z_L = 0.284\pm0.001$.

\section{Results}
\label{results}

\subsection{Metal Absorption}

We detect a large number of absorption lines from metal species in the high-resolution spectra at a redshift of $z_{\rm abs}=1.946938$ from various ionization stages: \ion{C}{i}, \ion{O}{i}, \ion{Si}{ii}, \ion{Fe}{ii}, \ion{S}{ii}, \ion{Ni}{ii}, \ion{Al}{ii}, \ion{Al}{iii}, \ion{C}{iv}, and \ion{Si}{iv}.

In order to obtain column densities of the singly ionized species (which is the dominant ionization state for metals in the neutral gas), we fit the transitions of \ion{Si}{ii}\,$\lambda\lambda1304,1808$, \ion{Fe}{ii}\,$\lambda\lambda1608, 1611$, \ion{S}{ii}\,$\lambda\lambda1250, 1253$ using Voigt profiles implemented in the Python module VoigtFit \citep{VoigtFit} with a total of 8 and 6 components for line of sight A and B, respectively.
We assume that the redshifts and Doppler broadening parameters of each components of the three heavy ions are identical. Thus the different $z$ and $b$ parameters for the components of S, Si and Fe are tied during the fit. In Table~\ref{tab:observables}, we give the total column densities derived for the two lines of sight. The best-fit parameters for the individual components are summarized in Appendix~\ref{app:metal_fit} together with figures showing the best-fit line profiles.

We furthermore fit the following absorption lines from \ion{C}{i} fine-structure transitions: \ion{C}{i}\,$\lambda1280$, $\lambda1560$ and $\lambda1656$. Since the individual velocity components are barely resolved, we assume that the excited fine-structure levels, $J=1$ (\ion{C}{i}$^*$) and $J=2$ (\ion{C}{i}$^{**}$), follow the same kinematic structure as the ground level ($J=0$), i.e., redshifts and broadening parameters for individual components are tied during the fit.
Moreover, due to the low signal-to-noise ratio, we assume a constant column density ratio of \ion{C}{i}$^*$ to \ion{C}{i} ($r^*$) and \ion{C}{i}$^{**}$ to \ion{C}{i} ($r^{**}$) for all components.
The total column densities derived for \ion{C}{i} and the column density ratios of the excited levels with respect to the ground level are summarized in Table~\ref{tab:observables}.
We identify two weak, intervening \ion{C}{iv} absorbers towards sightline A and one weak, intervening \ion{C}{iv} absorber towards sightline B. These are overlapping with the \ion{C}{i} line complexes and were included in the fit in order to obtain a good fit.
The best-fit parameters for the \ion{C}{i} and \ion{C}{iv} components included in the models are summarized in Appendix~\ref{app:metal_fit} together with a figure showing the individual lines used to constrain the fits.

Moreover, we observe two \ion{C}{iv} absorption systems along the two lines of sight at redshifts
$z=2.5861$ (intrinsic to the quasar) and $z=2.1179$.

\subsection{Atomic and Molecular Hydrogen}
\begin{figure}
	\includegraphics[width=1.0\columnwidth]{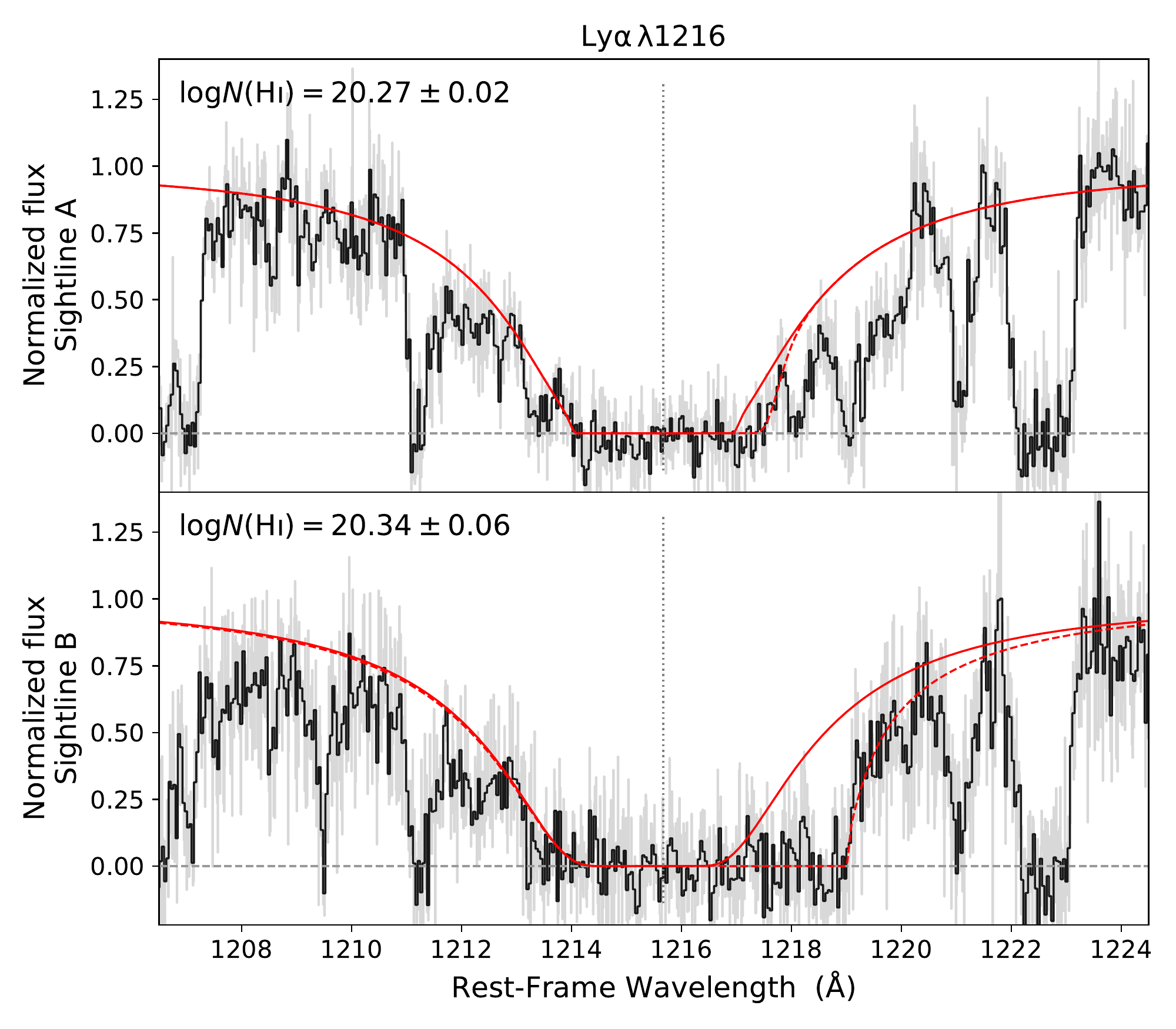}
	\caption{Comparison of the \lya\ absorption profiles for sightline A (top) and B (bottom)
	in the rest-frame of the absorber ($z_{\rm sys} = 1.946938$) from the Keck/HIRES spectra.
	The data have been rebinned for visual clarity (black line), and we show the unbinned data
	as the thin, grey line for comparison. The total best-fit profile is shown as the red, dashed line.
	The solid, red line shows the \ion{H}{i} profile for the bulk of the metal absorption.
	The dotted, vertical line marks the resonant wavelength in the systemic rest-frame.
	\label{fig:Lya}
	}
\end{figure}

We observe an asymmetry in the red wing of the \lya\ line profiles for both sightlines, yet more pronounced for sightline B, see Fig.~\ref{fig:Lya}. In order to properly model the absorption lines, we therefore fit a multi-component model to the \lya\ line. For both lines of sight, the first component is fixed to the average redshift of the low-ionization metals weighted by column-density. This `bulk' component properly reproduces the blue wing of the \lya\ profile in both sightlines.

The red wing is in turn fitted by adding a component to match the asymmetry. Similarly to the analysis by \citet{Noterdaeme2008}, the redder components require quite high Doppler broadening in order to fit the sharp edge of the red wing. For sightline B where the asymmetry is more pronounced, we need to include two red components in order to match the profile well. The best-fit column densities for the \ion{H}{i} component associated to the low-ionization metals are summarized in Table~\ref{tab:observables}. We do not observe any significant absorption from singly ionized species at the relative velocity of the redder components. However, the higher ionization lines, e.g., \ion{C}{iv} and \ion{Al}{iii}, do show absorption at similar relative velocities. We therefore argue that this gas phase has a higher ionization fraction and is not directly related to the gas carrying the bulk of the metals.

Moreover, we detect absorption lines from the Lyman-bands of H$_2$, ${\rm B}^1\Sigma_u^+(\nu) \leftarrow {\rm X}^1\Sigma_g^+(\nu=0)$ for $\nu$ up to $\nu=4$ for sightline A. Due to the more noisy data for sightline B, we only recover the first four bands up to $\nu=3$. We fit rotational levels up to $J=3$ for both sightlines, using one component for each $J$-level. Since the lines are damped, we cannot constrain the $b$-parameter. We therefore perform a first fit with an arbitrary, fixed $b$-parameter of $b=5$~km~s$^{-1}$ in order to obtain the redshift of the best-fit components. In both cases, the best-fit redshift corresponds to a component of \ion{C}{i} and we therefore tie the redshift and $b$-parameter to those of the matching \ion{C}{i} component for each sightline.
This agreement between \ion{C}{i} and H$_2$ components is consistent with the tight observed relationship between \ion{C}{i} and H$_2$ \citep{Noterdaeme2018}. We assume the same $b$-parameter for all $J$-levels and neglect any additional thermal contribution to the broadening due to the low temperature observed ($T\sim100$~K), see Sect.~\ref{phys_conditions}.
The observed spectra together with the best-fit models are shown in Fig.~\ref{fig:H2} and the best-fit parameters are summarized in Appendix~\ref{app:metal_fit}.
The molecular bands are highly blended with \lya\ forest and the low signal-to-noise ratio makes it very difficult to constrain all the molecular transitions, especially for sightline B. The tightest constraints to the fit come from the $\nu=0$ and $\nu=1$ bands.

Lastly, since the data are relatively noisy the fit has been performed independently by two individuals of our team using two different fitting software packages, and the two results are consistent within the rather large uncertainty ($\sim0.1-0.2$~dex).

\begin{figure*}
	\includegraphics[width=1.0\textwidth]{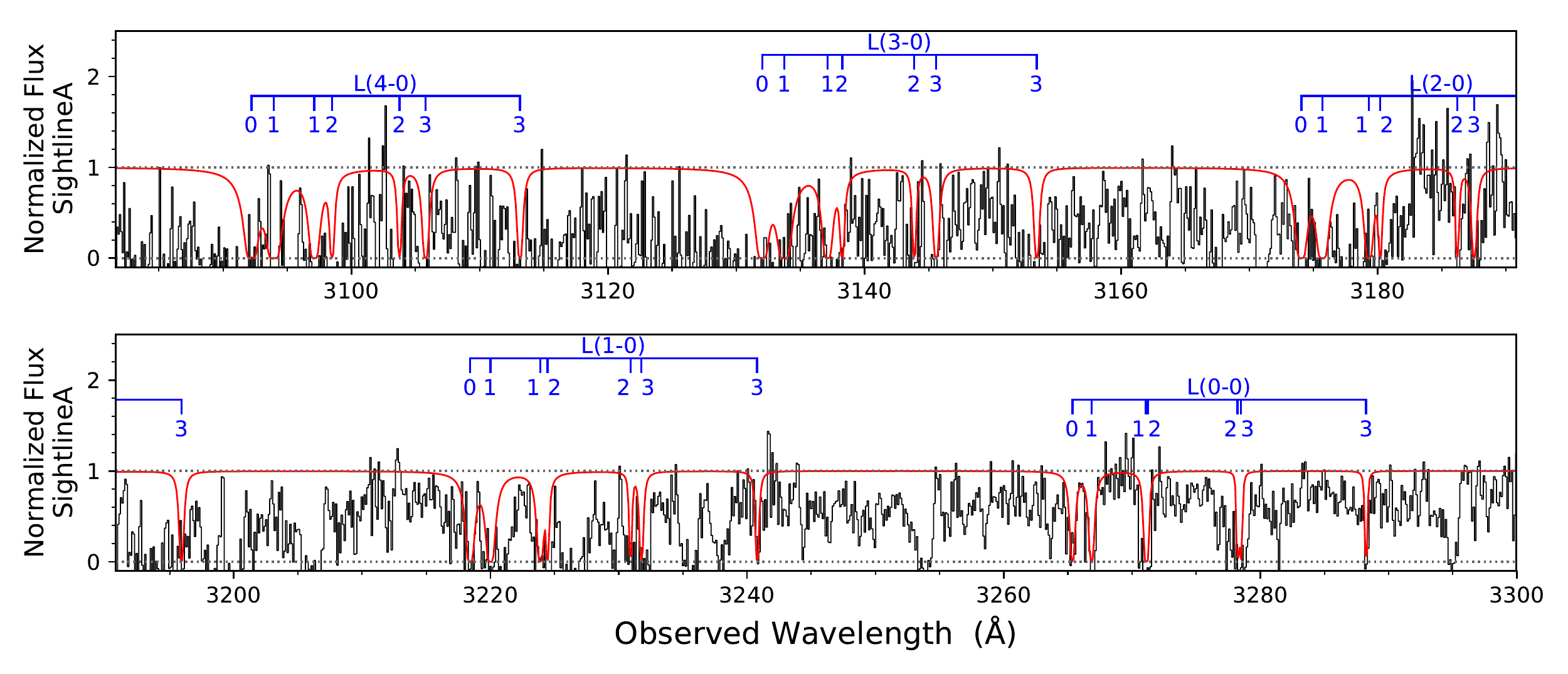}
	\includegraphics[width=1.015\textwidth]{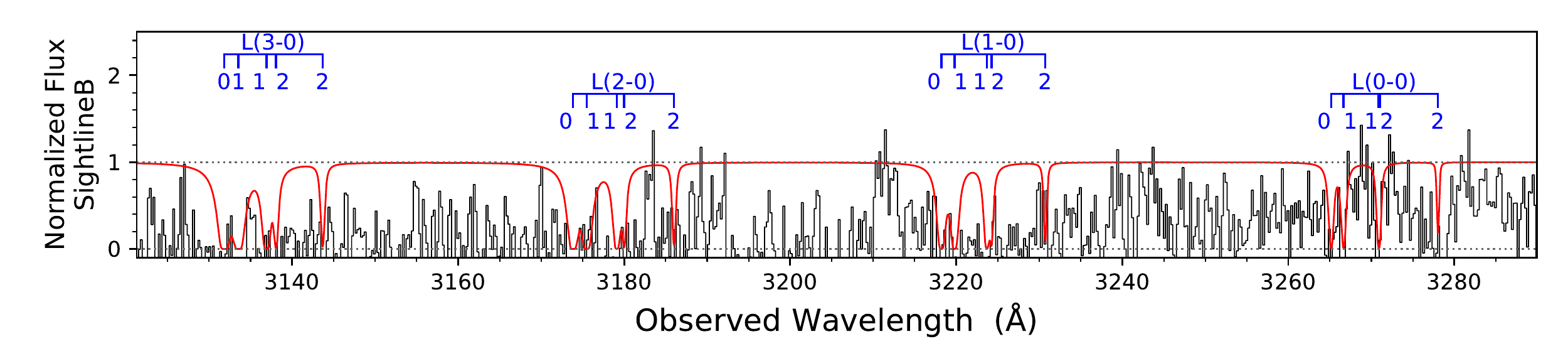}
	\caption{Molecular absorption bands of H$_2$ for sightline A (top two panels) and
	B (bottom panel) from the Keck/HIRES spectra. The locations of the rotational
	$J$-level for each of the vibrational bands are shown in blue above the spectra.
	The best-fit absorption profiles are shown in red. The data have been rebinned and smoothed
	(using Gaussian smoothing with a FWHM of 1.5 pixels) for visual clarity.
	\label{fig:H2}
	}
\end{figure*}

\subsection{Metallicity}
\label{metallicity}
Using the total hydrogen column density $N({\rm H}) = N({\rm H\, \textsc{i}}) + 2 N({\rm H_2})$, $\log N({\rm H}) = 20.4\pm0.1$ and $20.6\pm0.1$ for sightline A and B, respectively, we derive an average observed metallicity for the two lines of sight using sulphur since this element does not significantly deplete into dust grains \citep{Savage1996}. For sightlines A and B we find $[{\rm S/H}]_A = 0.0\pm0.1$ and $[{\rm S/H}]_B = -0.1\pm0.1$.

Since the observed column density ratio of S to Fe indicates significant amounts of dust in the gas, we have to consider dust corrections to the metallicities\footnote{We note that nucleosynthetic effects may play a role in the [S/Fe] ratio, however, at solar metallicity the enhancement of $\alpha$-elements is likely negligible \citep[e.g.,][]{Caffau2005, Berg2015}.}.
We use the depletion sequences as determined by \citet{DeCia2016} in order to calculate the fraction of metals locked up in dust grains. The formalism by \citet{DeCia2016} uses [Zn/Fe] as the overall tracer of the depletion sequences, however, as our high resolution spectra do not cover the zinc lines at $\lambda 2026$ and $\lambda 2062$, we use [S/Fe] as a tracer of [Zn/Fe] since both elements deplete very little into dust. Moreover, \citet{Berg2015} report [S/Zn]~$\approx0$ for high-metallicity DLAs \citep[see also][]{Rafelski2012, DeCia2016}. Using eq. (5) from the work by \citet{DeCia2016}, we obtain consistent dust-corrected metallicities for both sightlines, using both iron and sulphur, of $[{\rm Fe/H}]_0 = 0.2\pm0.1$ and $[{\rm S/H}]_0 = 0.3\pm0.1$, respectively.

Using the depletion sequences of \citet{DeCia2016}, we can furthermore calculate the expected optical extinction A(V) based on the observed metallicity and hydrogen column density.
For the two sightlines A and B, we then infer a rest-frame A(V) of $0.2\pm 0.1$ and $0.4\pm 0.1$, respectively. We note that for the calculation of A(V), we have used the {\it total} hydrogen column density, $\log N({\rm H})$, instead of just the atomic hydrogen column density $\log N({\rm H\,\textsc{i}})$ which is used in eq. (8) by \citet{DeCia2016}.

\begin{table}
  \caption{Overview of measurements \label{tab:observables}}
  \centering
  \begin{tabular}{lcc}
    \hline\vspace{2pt}
    Observable  &  LOS-A  &  LOS-B \\[2pt]
    \hline
    log N(\ion{H}{i})	& $20.27\pm0.02$  & $20.34\pm0.05$ \\
    log N(H$_2$)		&  $19.7\pm0.1$   &  $19.9\pm0.2$  \\
    log N(H)			&  $20.4\pm0.1$  &  $20.6\pm0.1$  \\
    $f_{\rm H_2}$		&  $0.3\pm0.1$  &   $0.4\pm0.1$  \\
    log N(Fe)			& $14.93\pm0.03$  & $14.80\pm0.03$ \\
    log N(Si)			& $15.55\pm0.03$  & $15.37\pm0.06$ \\
    log N(S)			& $15.52\pm0.03$  & $15.63\pm0.07$ \\
    $[{\rm Fe/S}]$		& $-1.0\pm0.1$  & $-1.2\pm0.1$ \\
    $[{\rm Si/S}]$		& $-0.4\pm0.1$  & $-0.7\pm0.1$ \\
    $A_V$				&  $0.19\pm0.02$  &  $0.36\pm0.02$ \\
    log N(\ion{C}{i})	& $14.25\pm0.02$  &  $14.79\pm0.05$  \\
    $\log r^{*}$		& $-0.40\pm0.04$  &  $-1.18\pm0.13$  \\
    $\log r^{**}$		& $-0.72\pm0.07$  &  $-1.65\pm0.14$  \\[2pt]
    \hline
  \end{tabular}
  
  \vspace{4mm}
  All column densities are given in units of cm$^{-2}$.
  
\end{table}

\section{Discussion}
\label{discussion}

\subsection{Physical Properties}
\label{phys_conditions}
Using the derived column densities of H$_2$ for the two lowest rotational levels ($J=0$ and $J=1$),
we can infer the excitation temperature, $T_{01}$, which is a good proxy for the overall kinetic
temperature of the molecular medium \citep*{Roy2006}.
For both sightlines we infer consistent temperatures of $T^A_{01} = 109\pm20$~K and $T^B_{01} = 89\pm25$~K.

\begin{figure}
	\includegraphics[width=1.0\columnwidth]{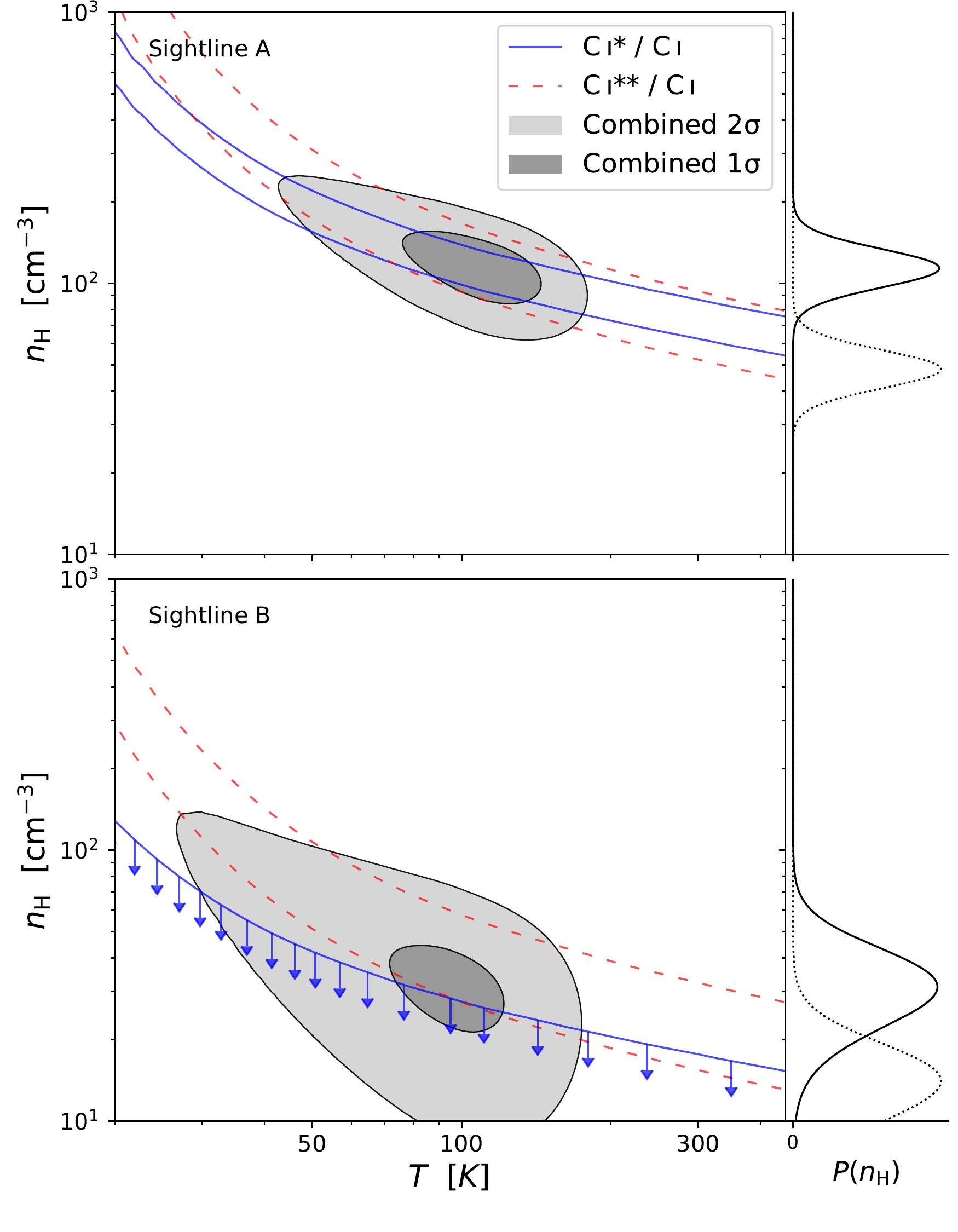}
	\caption{Density versus temperature derived from the observed excited \ion{C}{i} fine-structure levels
	in the absorber towards image A (top) and image B (bottom) assuming $f_{\rm H_2}=1$.
	The allowed $1\sigma$ ranges calculated for \ion{C}{i}$^*$ and \ion{C}{i}$^{**}$ are shown as the
	solid and dashed contours, respectively.
	The combined constraint on density and temperature is shown as the grey shaded regions.
	The curves on the right-hand axes show the marginalized probability density distributions
	for $n_{\rm H}$ assuming fully molecular gas (solid) or the observed, average molecular fraction (dotted).
	\label{fig:CI_density}
	}
\end{figure}

The \ion{C}{i} fine-structure levels can be excited either by collisions (in high-density or shielded gas)
or by UV pumping from the incident UV flux (in low-density gas). However, the incident radiation field
required to excite \ion{C}{i} to the observed levels for J1442+4055 is roughly 30 times higher than the Galactic UV field \citep{Habing1968}. We therefore neglect any contribution from the incident UV field in the following calculations, since the high amount of dust extinction indicates that the gas is efficiently shielded from UV photons\footnote{We caution that the dust reddening might not be fully associated with the cold gas phase.}.
We can then constrain the density of the cold gas phase assuming the temperature measurements derived from H$_2$ above.
We use the code {\sc Popratio} \citep{SilvaViegas2001} to calculate the expected population of the excited
fine-structure levels compared to the ground-state population in a grid of temperature and hydrogen density, $n_{\rm H}$.
For this calculation, we include a contribution from the cosmic microwave background at the absorber redshift
and from the extragalactic UV background field \citep{Khaire2018}\footnote{We have implemented the updated
calculation of the extragalactic UV background from \citet{Khaire2018} in {\sc Popratio} instead of the
default background field by \citet{Madau1999}.}.
Since most of the neutral hydrogen is likely not directly associated with the cold gas from which the bulk
of the \ion{C}{i} absorption arises, we argue that the molecular fraction in the cold gas phase is higher than
the average $f_{\rm H_2}$ and probably closer to unity, i.e., the \ion{C}{i} absorption arises predominantly
from fully molecular gas \citep[e.g.,][]{Bialy2016}.
Hence, we run the {\sc Popratio} models assuming $f_{\rm H_2} = 1$. The constraints from the observed
ratios of $N(\rm C\,\textsc{i}^*)/N(\rm C\,\textsc{i})$ and $N(\rm C\,\textsc{i}^{**})/N(\rm C\,\textsc{i})$
are shown in Fig.~\ref{fig:CI_density}. We subsequently take into account the temperature constraints from H$_2$
excitation (see shaded regions in Fig.~\ref{fig:CI_density}) and marginalize the 2-dimensional probability distribution
over $T$ in order to obtain the most probable density (right-hand panels of Fig.~\ref{fig:CI_density}).
The estimated hydrogen densities for sightlines A and B are $n_{\rm H}^{\textsc{a}}\approx110$~cm$^{-3}$ and $n_{\rm H}^\textsc{b}\approx40$~cm$^{-3}$.

For comparison, we run a set of models using the average, observed $f_{\rm H_2}$ integrated over
the full line of sight (see Table~\ref{tab:observables}).
For the lower molecular fraction, we obtain lower densities by a factor of 2.2
(see dotted lines in right-hand panels of Fig.~\ref{fig:CI_density}).
However, we caution that these estimates are strict lower limits, as the molecular
fraction inside the cold medium is certainly larger than the average integrated molecular fraction.

\subsection{Kinematics}

\begin{figure*}
	\includegraphics[width=1.0\textwidth]{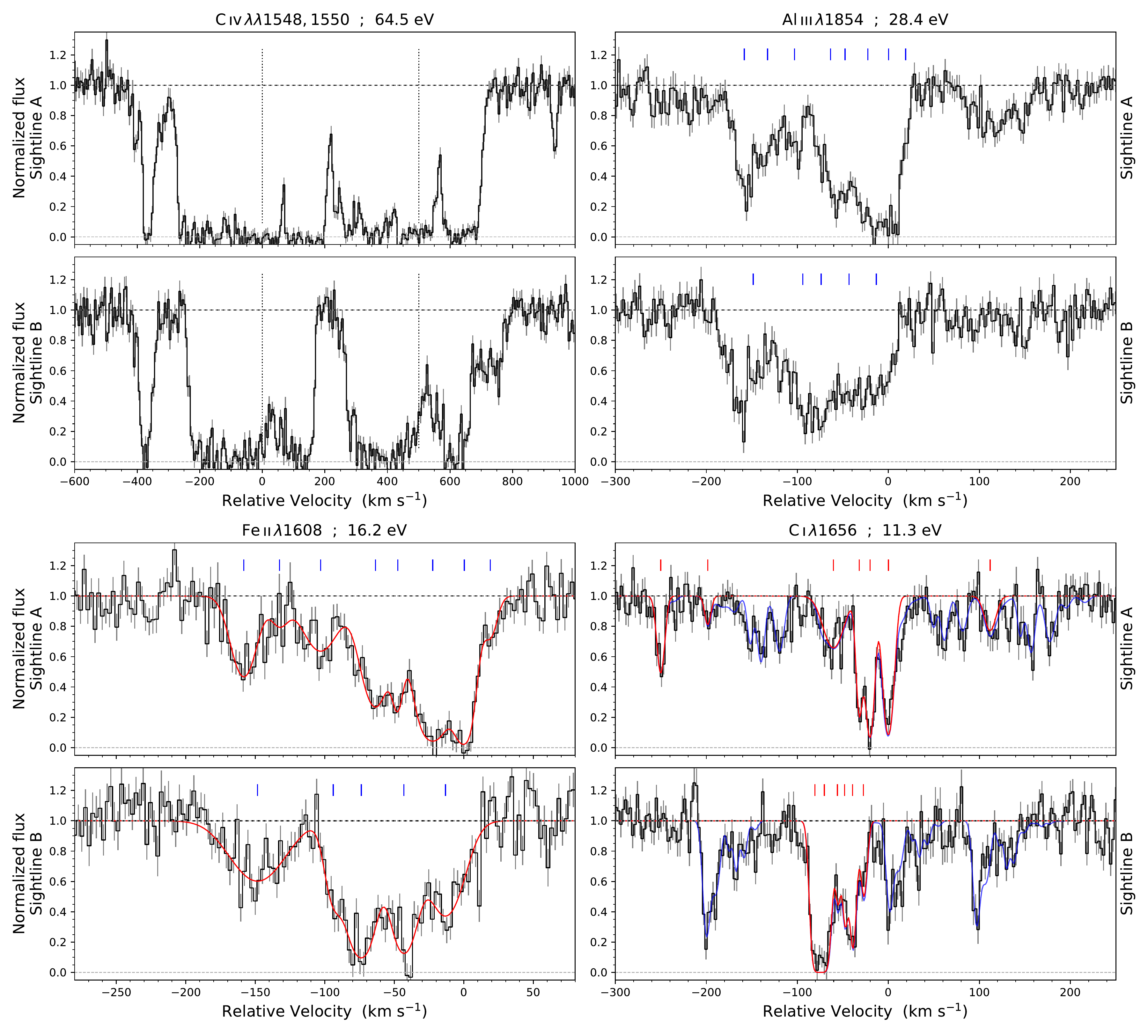}
	\caption{Comparison of lines of various ionization states
	(the ionization potential is given for each ion) for the
	two sightlines. All panels show velocities relative to
	$z_{\rm sys}=1.946938$.
	For \ion{C}{i} we show the total best-fit profile in blue,
	and highlight the absorption from ground state level in red
	(the red tick marks above show the location of the ground state components).
	For \ion{Fe}{ii}, we show the best-fit profile in red and
	components as blue tick marks.
	For comparison, the same velocity components are shown for \ion{Al}{iii}.
	For \ion{C}{iv} doublet, we show only the zero-velocities as dotted,
	vertical lines. The data have been rebinned by a factor of 2 for visual clarity.
	\label{fig:comparison}
	}
\end{figure*}

The low-ionization species (e.g., \ion{Si}{ii} and \ion{Fe}{ii}) for the two different lines of sight
show absorption over very similar velocity spreads $\Delta v \sim200$~km~s$^{-1}$ with several sub-components.
The individual components show different strengths, however, with the absorption being
dominated by a strong red component for sightline A defined as the systemic redshift.
In contrast, sightline B exhibits stronger absorption in the blueshifted component at
$v_{\rm rel} = -100$~km~s$^{-1}$. A similar pattern is observed in the higher ionization lines
of \ion{Al}{iii}. While the neutral carbon absorption is concentrated over a smaller range
in velocity space than the low-ionization species, the main component of \ion{C}{i} absorption
corresponds to the strongest \ion{Fe}{ii} component for both lines of sight (see Fig.~\ref{fig:comparison}).
There is a significant velocity shift for \ion{C}{i} between the two sightlines of
$\Delta v_{C\,\textsc{i}} \approx 100$~km~s$^{-1}$. This fits well with a scenario in which
the low-ionization species predominantly arise in a more warm diffuse phase (similar to
the warm neutral medium of the Milky Way) spread over a larger velocity width.
The cold gas clumps are more confined in velocity space likely tracing a more spatially confined structure as well.
If we interpret this velocity offset as a consequence of
ordered rotation, we can use the observed velocity shift as an indicator of the dynamical
mass of the system.
Since we do not directly observe the absorbing galaxy behind the lensing galaxy,
we have no prior information about the physical impact parameters of each line of sight through the absorbing galaxy. 
We therefore make the simplifying assumption that the two sightlines pierce symmetrically through
the galaxy at equal distances to the centre.
This allows us to infer a lower limit to the dynamical mass of $M_{\rm dyn} \gtrsim 5 \times 10^8~M_{\odot}$.

This is similar to the case of molecular absorption detected in a lensing galaxy
at $z=0.76$ towards the quasar PMN 0134$-$0931 presented by \citet{Wiklind2018}.
The authors find that the molecular gas
in the different lines of sight show velocity offsets of the order $\sim200$~km~s$^{-1}$.
Moreover, the lower limit to the dynamical mass is consistent with measurements of
stellar masses of DLAs at similar redshifts \citep{Christensen2014}, which similarly
serve as lower limits to the total dynamical mass.

\subsection{Filling Factor}

Expanding on the scenario laid out above, we can constrain the volume filling factor of the
cold gas. We assume that the cold gas must be distributed over scales of at least one kpc,
in order to observe \ion{C}{i} absorption in both sightlines. This is consistent with the findings
of \citet{Wiklind2018} who find molecular absorption spread out over 5~kpc with
a more or less uniform coverage \citep[see also][]{Carilli2013}.
In the following, we will estimate the volume filling factor,
assuming that the separation between the two sightlines is a representative scale over which
the cold gas is found. This will result in an upper limit to the filling factor, as the cold gas
might be spread over larger scales. Nonetheless, it is an instructive calculation
which puts direct constraints on the presence of cold gas in high-redshift galaxies.

We calculate the volume filling factor as the ratio:
\begin{equation}
	f_{\rm vol} = \frac{N v}{V}~,
\end{equation}

where $N$ is the total number of clouds in the volume $V$ and $v$ is the volume of each
cloud. We can also write $v$ as the product of the cross-section, $\sigma$, and typical
absorption length, $l$.
Alternatively, we can write the average number of detected clouds along a line of sight as:
\begin{equation}
	\mathcal{N} = \sigma L \frac{N}{V}~,
\end{equation}

where $L$ is the total absorption length along the line of sight through the medium.
Substituting the unknown cloud number density ($N/V$) in eq. (3) using $\mathcal{N}$, we arrive at:
\begin{equation}
	f_{\rm vol} = \mathcal{N} \frac{l}{L}~.
\end{equation}

In order to estimate the individual cloud absorption length, $l$, we can use the physical density derived above:

\begin{equation}
	l = \frac{ N_{H_2} }{ n_{\rm H_2} \mathcal{N}} = \frac{ 2 N_{\rm H_2} }{ n_{\rm H} \mathcal{N}} ~.
\end{equation}

Here we assume that the bulk of the molecular absorption arises from the cold gas phase
where the molecular fraction is close to unity, hence $n_{\rm H_2} = n_{\rm H} / 2$.
We derive similar cloud sizes (to within a factor of two) for both lines of sight: $l \approx 0.1$~pc.
This further strengthens our assumption that the conditions are similar for the various cold clouds.

We neglect geometrical effects and use the physical separation of the two sightlines at the redshift of the absorber,
$d_{\rm abs} = 0.7$~kpc, as an estimate of the absorption length along the line of sight, since there is no reason to
assume that the absorbing medium should be preferentially elongated along the line of sight nor perpendicular to it.
Thus we assume $L \geq d_{\rm abs}$. Note that this is a strict lower limit, as the \ion{C}{i}-bearing gas is most likely
distributed over larger scales both along the line of sight and perpendicular to it.

We use the number of observed velocity components in the \ion{C}{i} profile as an estimate of the number of cold clouds.
However, the number might be larger if some components are not resolved.
We find an average of $\mathcal{N} = 5$ cold clumps for both lines of sight, which translates to a volume filling factor of $f_{\rm vol} < 0.002$.
This is lower by almost an order of magnitude than the typical values of the CNM in the Milky Way ISM \citep[e.g.,][]{Ferriere2001}. 

Even though the volume filling factor is small, the cold gas is spread out over kpc scales, and hence the projected covering fraction, i.e., the probability of intersecting the cold gas, becomes large as evidenced by the detection of significant cold gas absorption in both lines of sight separated by $\sim$1~kpc.

While the covering fraction of cold, neutral gas may be high in the central parts of high-metallicity systems (where the pressure is high enough to allow efficient cooling), the bulk of the neutral gas in the overall population is dominated by the diffuse and warm gas spread over larger physical scales, as only $\sim1$~\% of \ion{H}{i} absorbers show significant \ion{C}{i} absorption \citep{Ledoux2015}. We can illustrate this scenario by a simple model in which the cross-section of DLAs and \ion{C}{i} absorbers is uniform within some typical physical scale. For DLAs at $z\sim2$, this typical scale is $R_{\rm H\textsc{i}}\sim10$~kpc but may extend up to $\sim30$~kpc or larger in a few cases \citep{Rubin2015}. Consequently, in order to match the 100 times lower incidence of \ion{C}{i} absorbers, the typical scale over which \ion{C}{i} bearing gas is detected must be $\sim10$ times smaller. Hence, $R_{\rm C\textsc{i}}$ is of the order a few kpc. This typical scale for cold gas absorption of a few kpc is in good agreement with the constraint inferred from the observations presented here, i.e., $R_{\rm C\textsc{i}} > 1$~kpc.
Taking into account the observed increase in \ion{C}{i} incidence towards lower redshift \citep{Ledoux2015}, we qualitatively match the larger $R_{\rm C\textsc{i}}\gtrsim 5$~kpc inferred at intermediate redshifts of $z=0.76$ \citep{Wiklind2018}.

It is important to keep in mind that \ion{C}{i} as a tracer of the cold gas is biased towards high-metallicity systems, due to the high abundance of carbon and due to more efficient cooling and dust shielding in high-metallicity gas.
Therefore, the incidence of cold gas is likely higher than that inferred purely by the incidence of \ion{C}{i} absorption. This matches well the conclusion by \citet{Neeleman2015} and \citet{Balashev2018} who find that the fraction of DLAs with significant amounts of cold gas is around 5\% for a metallicity-unbiased sample.

The present study therefore suggests that the inferred volume filling factor of $\sim0.1$~\% reproduces well the observed incidence of cold gas assuming a rather uniform cross-section of cold gas on kpc scales immersed within a more wide-spread warm, neutral gas phase extended on scales up to $10-30$~kpc.
This will provide important constraints for future numerical simulations designed to resolve directly the cold neutral and diffuse molecular phases of the ISM in high-redshift galaxies.

\section{Summary}

	We here report the analysis of the $z_{\rm abs}=1.9469$ absorber seen
	towards both images of the lensed quasar J1442+4055 at $z_{\textsc{QSO}}=2.590$.
	We obtain imaging data from the Nordic Optical Telescope in the $r$-band
	and combine these data with archival SDSS imaging data in $u$, $g$, $r$,
	$i$, and $z$ bands. After subtracting the PSF contribution from the quasar
	images, we detect the lensing galaxy in all filters except $u$-band.
	The higher quality $r$-band image from the NOT allows us to constrain
	the structural parameters whereas the larger filter coverage of the SDSS
	photometry enables us to obtain a photometric redshift of
	$z_{\rm phot}=0.35\pm0.15$.

	We are able to recover the lensing galaxy spectrum by fitting the two
	LRIS spectra (covering wavelengths from 3600$-$8600~\AA) with a combined
	quasar and galaxy model. The contribution of the lensing galaxy flux
	in each spectrum of the quasar was constrained from the NOT imaging
	data taken during similar conditions as the spectra.
	The recovered lens galaxy spectrum suffers from artefacts due to variations
	in the quasar emission lines and continuum. However, we clearly detect
	the 4000~\AA\ break together with \ion{Ca}{ii}, \ion{Mg}{i}, and
	\ion{Na}{i} absorption lines.
	The best-fit redshift of the lensing galaxy is $z_L = 0.284\pm0.001$.
	Using the lens redshift, we obtain a physical line-of-sight separation
	at the absorber redshift of $d_{\rm abs} = 0.7$~kpc.
	
	We detect several metal species from different ionization stages in
	absorption from the $z_{\rm abs}=1.9469$ absorber. Moreover, we detect
	both neutral and molecular hydrogen absorption along both lines of sight.
	The inferred column densities of metals and hydrogen give consistent
	dust-corrected metallicities along both sightlines of $[{\rm S/H}] = 0.3\pm0.1$.
	The detection of strong absorption from H$_2$, \ion{C}{i} and its
	fine-structure levels  indicates the presence of cold gas along both
	lines of sight.
	
	The excitation of the lowest rotational levels of H$_2$ provides a
	measure of the kinetic temperature of the gas phase harbouring
	molecules and \ion{C}{i}.
	For both lines of sight, we infer consistent temperature estimates of
	$T_A=109\pm20$ and $T_B = 89\pm25$~K.
	Using the estimated temperatures together with the observed fine-structure
	levels of neutral carbon, we model the gas cloud density using {\sc Popratio}.
	The inferred average densities for the cold gas phase along the two lines
	of sight are $n_{\rm H}^{\textsc{a}}\approx110$~cm$^{-3}$
	and $n_{\rm H}^\textsc{b}\approx40$~cm$^{-3}$.
	
	We find that the low-ionization species (e.g., \ion{Fe}{ii} and \ion{Si}{ii})
	exhibit similar velocity widths with numerous components. The relative
	strength of the various components differ for each sightline. The neutral
	absorption lines correlate with the strongest component of the low-ionization
	species. The neutral absorption in sightline B exhibits a velocity offset
	of $\sim$100~km~s$^{-1}$ relative to the absorption in sightline A.
	If we interpret this velocity offset as a pure result of rotation in an
	ordered disk, we obtain a lower limit to the system's dynamical mass of
	$M_{\rm dyn} \gtrsim 5\times 10^8$~M$_{\odot}$.
	
	Using the physical densities to infer the typical size of cold gas clouds
	($l\sim$0.1~pc), we infer an upper limit to the volume filling factor of
	cold gas in this galaxy assuming that the cold gas is distributed over at
	least 0.7~kpc corresponding to the line-of-sight separation.
	This yields a volume filling factor of $f_{\rm vol} < 0.1$~\%.

	We argue that the observations are consistent with a picture in which cold
	gas (as probed by neutral carbon) is confined in small clouds (with sizes
	of sub-pc to a few pc) that are distributed over kpc-scales in high redshift
	galaxies. This cold gas phase of $T\sim 100$~K is immersed in a warmer
	($T\sim10^4$~K) gas phase which can in turn extend up to a few tens of kpc,
	with neutral hydrogen probing both these phases.
	This simple picture is able to explain the observed incidence rate of DLAs
	and \ion{C}{i} absorbers.

\begin{acknowledgements}

We wish to thank the anonymous referee whose constructive comments helped
improve the lens model significantly.
The research leading to these results has received funding from the French
{\sl Agence Nationale de la Recherche} under grant no ANR-17-CE31-0011-01
(project ``HIH2'' -- PI Noterdaeme).
MF acknowledges support by the Science and Technology Facilities Council 
[grant number ST/P000541/1]. This project has received funding from the 
European Research Council (ERC) under the European Union's Horizon 2020 
research and innovation programme (grant agreement No 757535).
The Cosmic Dawn Center is funded by the Danish National Research Foundation.
S.B. is supported by RSF grant 18-12-00301.
F. C. acknowledges support from the Swiss National Science Foundation (SNSF).
M.R. acknowledges support by a NASA Keck PI Data Award, administered by the NASA Exoplanet Science Institute. 
Based on observations made with the Nordic Optical Telescope, operated on the island of La Palma jointly by Denmark, Finland, Iceland, Norway, and Sweden, in the Spanish Observatorio del Roque de los Muchachos of the Instituto de Astrof\'isica de Canarias.
Some of the data presented herein were obtained at the W. M. Keck Observatory from telescope time allocated to the National Aeronautics and Space Administration through the agency's scientific partnership with the California Institute of Technology and the University of California. The Observatory was made possible by the generous financial support of the W. M. Keck Foundation. The authors wish to recognize and acknowledge the very significant cultural role and reverence that the summit of Maunakea has always had within the indigenous Hawaiian community. We are most fortunate to have the opportunity to conduct observations from this mountain.

\end{acknowledgements}

\bibliographystyle{aa}

\begin{appendix}

\section{Absorption Line Fits}
\label{app:metal_fit}

\begin{figure*}
	\includegraphics[width=1.0\textwidth]{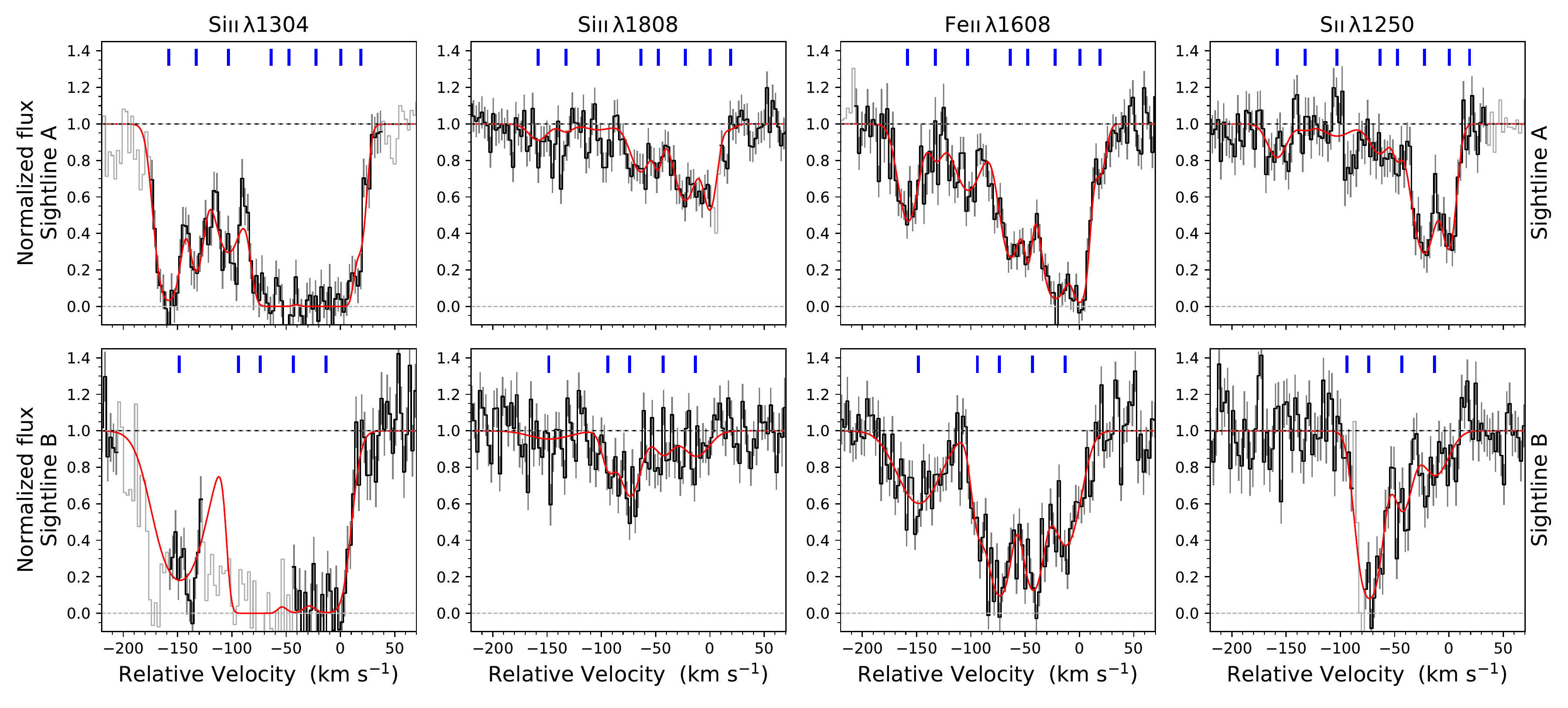}
	\caption{Comparison of singly ionized species for the two lines of sight.
	Regions of the spectra used to constrain the Voigt profile fitting are shown
	in black (gray regions were masked out during the fit due to \lya-forest contamination).
	The best-fit profiles are shown in red and the blue tick marks above the spectra
	indicate the positions of the individual components.
	\label{fig:metals}
	}
\end{figure*}

We here provide the best-fit parameters for the individual components of the Voigt profile models for low-ionization species (Tables~\ref{tab:metals_A} and \ref{tab:metals_B}). In Fig.~\ref{fig:metals}, we show the best-fit profiles of the low-ionization transitions.

\begin{figure*}
	\includegraphics[width=1.0\textwidth]{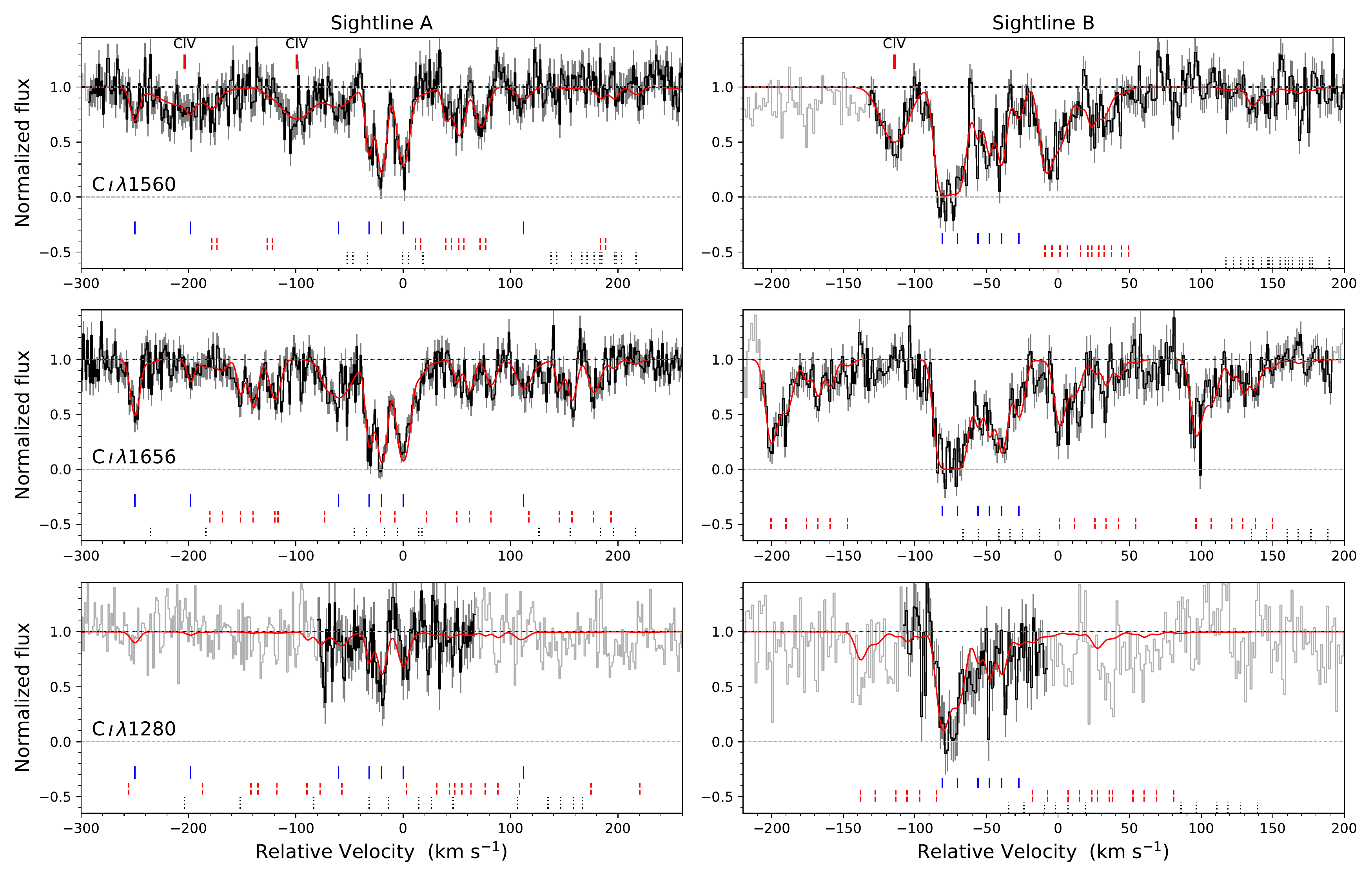}
	\caption{Comparison of \ion{C}{i} fine-structure complexes for the two lines of sight.
	Regions of the spectra used to constrain the model are shown
	in black (gray regions were masked out during the fit). The best-fit profiles
	are shown in red and the tick marks below the spectra indicate the positions
	of the individual components of the $J=0$ (blue solid), $J=1$ (red dashed),
	and $J=2$ (black dotted) transitions.
	The red tick marks above the spectra mark the location of weak, intervening
	\ion{C}{iv} systems.
	\label{fig:CI}
	}
\end{figure*}

The best-fit parameters for the various rotational levels of H$_2$ lines for the two lines of sight are given in Tables~\ref{tab:H2_A} and \ref{tab:H2_B}, and in Tables~\ref{tab:CI_A} and \ref{tab:CI_B}, we give the best-fit parameters for the \ion{C}{i} fine-structure lines for the two lines of sight. The best-fit line profiles for the \ion{C}{i} complexes used to constrain the model are shown in Figure~\ref{fig:CI}.

\begin{table*}
	\caption{Best-fit Voigt profile parameters for individual components of
	singly ionized metal species along sightline A
	\label{tab:metals_A}}
	\centering
	\begin{tabular}{ccccc}
	\hline\vspace{2pt}
	Rel. velocity\tablefootmark{a} &  $b$           & \multicolumn{3}{c}{$\log(N/{\rm cm^{-2}})$} \\
	(km~s$^{-1}$)     &  (km~s$^{-1}$) &    \ion{Fe}{ii}    &    \ion{S}{ii}   &      \ion{Si}{ii} \\[2pt]
	\hline
	$-158.3 \pm 0.8$  &  $11.3 \pm 1.1$  & $13.81 \pm 0.04$  & $14.38 \pm 0.11$  & $14.29 \pm 0.06$ \\
	$-132.6 \pm 1.1$  &  $ 7.7 \pm 1.8$  & $13.09 \pm 0.19$  & $13.45 \pm 0.52$  & $13.82 \pm 0.08$ \\
	$-103.0 \pm 1.2$  &  $16.7 \pm 2.7$  & $13.74 \pm 0.06$  & $14.07 \pm 0.13$  & $14.00 \pm 0.06$ \\
	$ -63.6 \pm 1.2$  &  $12.3 \pm 1.4$  & $14.08 \pm 0.05$  & $14.36 \pm 0.07$  & $14.84 \pm 0.06$ \\
	$ -47.5 \pm 0.8$  &  $ 3.8 \pm 1.6$  & $13.63 \pm 0.12$  & $13.90 \pm 0.20$  & $14.30 \pm 0.19$ \\
	$ -22.5 \pm 0.9$  &  $13.3 \pm 1.3$  & $14.49 \pm 0.05$  & $15.23 \pm 0.04$  & $15.12 \pm 0.05$ \\
	$   0.4 \pm 0.7$  &  $ 7.3 \pm 0.7$  & $14.34 \pm 0.08$  & $14.95 \pm 0.07$  & $14.94 \pm 0.07$ \\
	$  19.0 \pm 0.0$  &  $ 5.4 \pm 1.1$  & $13.17 \pm 0.10$  &     $<14.07$      & $13.53 \pm 0.08$ \\
	\hline
	\end{tabular}
	\tablefoot{
	\tablefoottext{a}{Relative to $z_{\rm sys} = 1.946938$}
	}
\end{table*}

\begin{table*}
	\caption{Best-fit Voigt profile parameters for individual components of
	singly ionized metal species along sightline B
	\label{tab:metals_B}}
	\centering
	\begin{tabular}{ccccc}
	\hline\vspace{2pt}
	Rel. velocity\tablefootmark{a} &  $b$           & \multicolumn{3}{c}{$\log(N/{\rm cm^{-2}})$} \\
	(km~s$^{-1}$)     &  (km~s$^{-1}$) &    \ion{Fe}{ii}     &   \ion{S}{ii}     &    \ion{Si}{ii} \\[2pt]
	\hline
	$-148.4 \pm 2.0$  &  $25.2 \pm 3.5$  & $13.97 \pm 0.07$  &        --         & $14.32 \pm 0.07$\\
	$ -94.0 \pm 0.0$\tablefootmark{b}  &  $ 7.4 \pm 1.9$  & $13.61 \pm 0.11$  &     $<11.78$      & $14.52 \pm 0.17$\\
	$ -73.9 \pm 0.9$  &  $12.0 \pm 1.4$  & $14.32 \pm 0.05$  & $15.50 \pm 0.08$  & $15.00 \pm 0.08$\\
	$ -43.2 \pm 1.1$  &  $10.7 \pm 1.8$  & $14.22 \pm 0.06$  & $14.81 \pm 0.10$  & $14.45 \pm 0.20$\\
	$ -13.2 \pm 1.9$  &  $15.8 \pm 1.9$  & $14.06 \pm 0.06$  & $14.66 \pm 0.12$  & $14.64 \pm 0.11$\\
	\hline
	\end{tabular}
	
	\tablefoot{
	\tablefoottext{a}{Relative to $z_{\rm sys} = 1.946938$.}
	\tablefoottext{b}{The redshift was kept fixed to the value obtained from a fit to \ion{Fe}{ii} only.}
	}
\end{table*}

\begin{table}
	\caption{Best-fit parameters for H$_2$ for sightline A
	\label{tab:H2_A}}
	\centering
	\begin{tabular}{cccc}
	\hline\vspace{2pt}
		 Rel. velocity\tablefootmark{a} &       $b$       &  Rot. level  &  $\log(N/{\rm cm^{-2}})$ \\
		   (km~s$^{-1}$)   &  (km~s$^{-1}$)  &              &                          \\[2pt]
	\hline
              $-19$      &      $4.2$      &  H$_2, J=0$  &  $19.2 \pm 0.1$  \\
               --        &        --       &  H$_2, J=1$  &  $19.4 \pm 0.1$  \\
               --        &        --       &  H$_2, J=2$  &  $18.4 \pm 0.3$  \\
               --        &        --       &  H$_2, J=3$  &  $18.7 \pm 0.2$  \\
	\hline
	\end{tabular}
	
	\tablefoot{
	\tablefoottext{a}{Relative to $z_{\rm sys} = 1.946938$}
	}
	
\end{table}

\begin{table}
	\caption{Best-fit parameters for H$_2$ for sightline B
	\label{tab:H2_B}}
	\centering
	\begin{tabular}{cccc}
	\hline\vspace{2pt}
		 Rel. velocity\tablefootmark{a} &       $b$       &  Rot. level  &  $\log(N/{\rm cm^{-2}})$ \\
		   (km~s$^{-1}$)   &  (km~s$^{-1}$)  &              &                          \\[2pt]
	\hline
              $-42$      &      $2.9$      &  H$_2, J=0$  &  $19.5 \pm 0.2$  \\
               --        &        --       &  H$_2, J=1$  &  $19.6 \pm 0.2$  \\
               --        &        --       &  H$_2, J=2$  &  $18.7 \pm 0.2$  \\
	\hline
	\end{tabular}
	
	\tablefoot{
	\tablefoottext{a}{Relative to $z_{\rm sys} = 1.946938$}
	}

\end{table}

\begin{table*}
	\caption{Best-fit parameters for \ion{C}{i} for sightline A
	\label{tab:CI_A}}
	\centering
	\begin{tabular}{ccccc}
	\hline\vspace{2pt}
	Rel. velocity\tablefootmark{a} &       $b$        &  \multicolumn{3}{c}{$\log(N/{\rm cm^{-2}})$\tablefootmark{b}}   \\
	  (km~s$^{-1}$)   &  (km~s$^{-1}$)   &        \ion{C}{i}        &       \ion{C}{i}$^*$    &       \ion{C}{i}$^{**}$ \\[2pt]
	\hline
	$-250.0 \pm 0.5$  &  $ 4.4 \pm 0.8$  &  $13.08 \pm 0.05$ & $12.68 \pm 0.06$ & $11.90 \pm 0.13$ \\
	$-198.4 \pm 1.3$  &  $ 3.0 \pm 2.9$  &  $12.46 \pm 0.14$ & $12.06 \pm 0.15$ & $11.28 \pm 0.19$ \\
	$ -60.3 \pm 1.4$  &  $14.4 \pm 1.9$  &  $13.25 \pm 0.05$ & $12.85 \pm 0.06$ & $12.06 \pm 0.13$ \\
	$ -31.9 \pm 0.3$  &  $ 2.5 \pm 0.2$  &  $13.48 \pm 0.06$ & $13.08 \pm 0.05$ & $12.30 \pm 0.14$ \\
	$ -20.1 \pm 0.3$  &  $ 4.2 \pm 0.4$  &  $13.75 \pm 0.04$ & $13.35 \pm 0.03$ & $12.57 \pm 0.13$ \\
	$   0.0 \pm 0.3$  &  $ 5.7 \pm 0.4$  &  $13.71 \pm 0.03$ & $13.31 \pm 0.03$ & $12.53 \pm 0.12$ \\
	$ 111.9 \pm 1.6$  &  $ 7.8 \pm 2.4$  &  $12.80 \pm 0.10$ & $12.40 \pm 0.10$ & $11.62 \pm 0.15$ \\[3pt]
	\hline
	$z$   &   &   \ion{C}{iv}   &   &   \\
	\hline
	$1.96408 \pm 0.00001$  & $19.7 \pm 1.6$ & $13.50 \pm 0.03$ &  &  \\
	$1.96800 \pm 0.00004$  & $31.2 \pm 6.2$ & $13.14 \pm 0.07$ &  &  \\
	\hline
	
	\end{tabular}
	
	\tablefoot{
	\tablefoottext{a}{Relative to $z_{\rm sys} = 1.946938$}
	\tablefoottext{b}{The column densities for \ion{C}{i}$^*$ and \ion{C}{i}$^{**}$ were tied to the ground level
		assuming two freely-varying column density ratios $r^*$ and $r^{**}$ for all components.}
	}

\end{table*}

\begin{table*}
	\caption{Best-fit parameters for \ion{C}{i} for sightline B
	\label{tab:CI_B}}
	\centering
	\begin{tabular}{ccccc}
	\hline\vspace{2pt}
	Rel. velocity\tablefootmark{a} &       $b$        &  \multicolumn{3}{c}{$\log(N/{\rm cm^{-2}})$\tablefootmark{b}}   \\
	  (km~s$^{-1}$)   &  (km~s$^{-1}$)   &        \ion{C}{i}        &       \ion{C}{i}$^*$    &       \ion{C}{i}$^{**}$ \\[2pt]
	\hline
	$ -80.7 \pm 0.5$  &  $ 3.6 \pm 0.4$  &  $14.47 \pm 0.09$ & $13.75 \pm 0.06$ & $12.82 \pm 0.13$ \\
	$ -70.3 \pm 1.0$  &  $ 5.0 \pm 0.9$  &  $14.16 \pm 0.09$ & $13.44 \pm 0.08$ & $12.51 \pm 0.14$ \\
	$ -55.9 \pm 0.6$  &  $ 1.1 \pm 0.2$  &  $13.54 \pm 0.15$ & $12.82 \pm 0.13$ & $11.90 \pm 0.19$ \\
	$ -48.1 \pm 0.4$  &  $ 1.3 \pm 0.2$  &  $13.88 \pm 0.12$ & $13.16 \pm 0.10$ & $12.23 \pm 0.15$ \\
	$ -39.3 \pm 0.4$  &  $ 2.6 \pm 0.1$  &  $13.67 \pm 0.07$ & $12.95 \pm 0.08$ & $12.02 \pm 0.14$ \\
	$ -27.4 \pm 0.7$  &  $ 3.0 \pm 0.0$  &  $13.01 \pm 0.08$ & $12.29 \pm 0.10$ & $11.37 \pm 0.15$ \\[3pt]
	\hline
	$z$   & $b$  &  $\log(N/{\rm cm^{-2}})$   &   &   \\
	      & (km~s$^{-1}$)  &   \ion{C}{iv}   &   &   \\
	\hline
	$1.963932 \pm 0.000006$ & $11.5 \pm 0.9$ & $13.58 \pm 0.03$ &  &  \\
	\hline
	\end{tabular}
	
	\tablefoot{
	\tablefoottext{a}{Relative to $z_{\rm sys} = 1.946938$}
	\tablefoottext{b}{The column densities for \ion{C}{i}$^*$ and \ion{C}{i}$^{**}$ were tied to the ground level
		assuming two freely-varying column density ratios $r^*$ and $r^{**}$ for all components.}
	}
		
\end{table*}

\end{appendix}

\end{document}